\documentclass[twocolumn,tighten,onecolappendix]{aastex701}

\usepackage{amssymb,amsmath,amsthm}
\usepackage{graphicx}
\usepackage{xspace}
\usepackage{mathtools}

\def\uas{\ensuremath{\mu}as\xspace} 
\def\msun{M\ensuremath{_{\odot}}\xspace} 
\def\m87{M87$^*$\xspace} 
\def\sgra{Sgr\,A$^*$\xspace} 

\defcitealias{M87Paper1}{EHTC et al. (2019a)}
\defcitealias{M87Paper2}{EHTC et al. (2019b)}
\defcitealias{M87Paper3}{EHTC et al. (2019c)}
\defcitealias{M87Paper4}{EHTC et al. (2019d)}
\defcitealias{M87Paper5}{EHTC et al. (2019e)}
\defcitealias{M87Paper6}{EHTC et al. (2019f)}
\defcitealias{M87Paper7}{\m87~Paper~7}
\defcitealias{M87Paper8}{\m87~Paper~8}
\defcitealias{SgrA_Paper1}{\sgra~Paper~1}
\defcitealias{SgrA_Paper2}{\sgra~Paper~2}
\defcitealias{SgrA_Paper3}{\sgra~Paper~3}
\defcitealias{SgrA_Paper4}{\sgra~Paper~4}
\defcitealias{SgrA_Paper5}{\sgra~Paper~5}
\defcitealias{SgrA_Paper6}{\sgra~Paper~6}

\makeatletter
\DeclareRobustCommand{\pder}[1]{%
  \@ifnextchar\bgroup{\@pder{#1}}{\@pder{}{#1}}}
\newcommand{\@pder}[2]{\frac{\partial#1}{\partial#2}}
\makeatother

\begin{document}

\title{A model for the image structure and spectral energy distribution of relativistic jets}

\correspondingauthor{Dominic~W.~Pesce}

\author[0000-0002-5278-9221]{Dominic~W.~Pesce}
\affiliation{Center for Astrophysics $|$ Harvard \& Smithsonian, 60 Garden Street, Cambridge, MA 02138, USA}
\affiliation{Black Hole Initiative at Harvard University, 20 Garden Street, Cambridge, MA 02138, USA}
\email[show]{dpesce@cfa.harvard.edu}

\author[0000-0002-1919-2730]{Ramesh~Narayan}
\affiliation{Center for Astrophysics $|$ Harvard \& Smithsonian, 60 Garden Street, Cambridge, MA 02138, USA}
\affiliation{Black Hole Initiative at Harvard University, 20 Garden Street, Cambridge, MA 02138, USA}
\email{rnarayan@cfa.harvard.edu}

\author[0000-0003-0213-7628]{Yuh~Tsunetoe}
\affiliation{Shanghai Astronomical Observatory, Chinese Academy of Sciences, Shanghai 200030, P. R. China}
\affiliation{Center for Computational Sciences, University of Tsukuba, 1-1-1 Tennodai, Tsukuba, Ibaraki 305-8577, Japan}
\email{ytsunetoe@ccs.tsukuba.ac.jp}

\begin{abstract}
Upcoming advances in high-frequency VLBI promise to enable spatially resolved studies of astrophysical jets across an increasingly large source population.  We present here a Python implementation of the semi-analytic model from \citet{Tsunetoe_2025} for self-consistently interpreting the spatial and spectral structure of the synchrotron emission from such jets.  The model is based on an axisymmetric force-free jet solution supplemented with a prescription for nonthermal electron heating, synchrotron cooling, pitch-angle anisotropy in the electron distribution, and radiative transfer along straight rays.  We use this model to explore how the predicted spectral energy distribution depends on the model parameters, identifying approximately self-similar scalings with black hole mass, accretion rate, and electron heating efficiency.  We also compare the model predictions against observations of \m87 and \sgra.  For \m87, a single fiducial parameter set qualitatively reproduces VLBI images across multiple frequencies and matches the observed longitudinal intensity profile, while also yielding a broadband SED that is quantitatively consistent with the observed radio through X-ray measurements.  For \sgra, a fit to the quiescent-state SED demonstrates that emission from a collimated jet with comparable nonthermal heating efficiency to \m87's is quantitatively consistent with the observed radio through X-ray measurements, predicting a compact jet whose radio size decreases inversely with frequency, consistent with VLBI constraints.
\end{abstract}

\section{Introduction}\label{sec:Intro}

Relativistic astrophysical jets provided some of the earliest evidence for the existence of black holes, and they remain one of the clearest observational signatures of supermassive black hole (SMBH) ubiquity.  The canonical picture for these jets holds that magnetic fields threading the event horizon of a spinning black hole extract energy and angular momentum from the hole, powering an outflow containing relativistic particles that we observe via their synchrotron emission \citep{Blandford_1977,Blandford_1978,Blandford_2019}.  Because of the high brightness temperature of this synchrotron emission, relativistic jets have been the target of very long baseline interferometric (VLBI) radio observations since the development of the technique \citep{Bare_1967,Broten_1967}.  Jetted active galactic nuclei (AGN) were the first sources to be imaged with VLBI \citep{Readhead_1979,Pearson_1981}, and they remain among the most frequently observed VLBI targets.

One of the most well-known and spectacular examples of relativistic jets is found in the M87 galaxy and associated with the \m87 black hole, whose proximity and large mass have made it the target of increasingly high-resolution VLBI observations for decades \citep[e.g.,][]{Reid_1989,Junor_1999,Asada_2012,Walker_2018,Lu_2023}.  The \m87 jet has now been imaged with VLBI over a large dynamic range in spatial scales, revealing a broad opening angle near the base, an approximately parabolic collimation profile, and a persistent limb-brightened morphology \citep{Kim_2018,Nikonov_2023}.  However, \m87 is an exceptional case; for the majority of jetted systems, VLBI primarily provides only longitudinal information such as core shifts and brightness temperature profiles.  The transverse structure -- and particularly the structure of the innermost regions of the jet -- is only resolved in a relatively small number of nearby systems \citep[e.g.,][]{Boccardi_2017,Kovalev_2020}.

That situation is changing as (sub)millimeter VLBI capabilities grow, providing more routine access to ever-finer angular scales.  The Event Horizon Telescope (EHT) has already observed multiple AGN jets \citep{Kim_2020,Janssen_2021,Issaoun_2022,Jorstad_2023,Roeder_2025}, and this trend is set to accelerate substantially in the coming years as the EHT continues to expand its capabilities \citep{EHT_2024}.  The next-generation EHT (ngEHT) project will improve the EHT's sensitivity through increased bandwidths and the addition of multiple new telescopes to the array \citep{Doeleman_2023,Johnson_2023}, while also implementing simultaneous multi-frequency observing capabilities to push the angular resolution frontier by enabling routine observations at the highest observing frequencies \citep{Issaoun_2023,Rioja_2023}.  The Black Hole Explorer (BHEX) mission will sharpen the angular resolution of the array even further via the addition of a spaceborne element \citep{Johnson_2024}.  These upgrades will bring a population of partially and fully resolved jetted sources into view \citep{Ramakrishnan_2023,Lico_2023,Kawashima_2024,Nair_2024}, shifting the science focus from individual case studies to demographic considerations.

A rich collection of tools exists for simulating the emission from AGN jets.  General relativistic magnetohydrodynamics (GRMHD) codes \citep[e.g.,][]{Gammie_2003,Porth_2017,Liska_2018,Prather_2021} and ray-tracing and radiative transfer pipelines \citep[e.g.,][]{Noble_2007,Dolence_2009,Dexter_2016,Pu_2016,Bronzwaer_2018,Moscibrodzka_2018} provide general-purpose physics packages that cover important parts of the jet emission problem; such tools are widely used for, e.g., interpretation of the near-horizon emission observed by the EHT \citep{Porth_2019,Gold_2020,Wong_2022,Prather_2023}.  Special relativistic MHD codes such as \texttt{PLUTO} \citep{Mignone_2007} have likewise been extensively used to simulate the dynamics and structure of relativistic jets on larger scales (see, e.g., \citealt{Komissarov_2021}, for a review).  More jet-specific tools also exist, including broadband spectral energy distribution (SED) models \citep[e.g.,][]{Lucchini_2022} and both semi-analytic as well as phenomenological image-plane models \citep[e.g.,][]{Broderick_2009,Chang_2024,GomezMiller_2026}.

However, the existing jet modeling capabilities are largely distributed across specialized workflows.  GRMHD codes and radiative transfer pipelines typically require substantial expertise and nontrivial post-processing of simulation products, while semi-analytic or phenomenological packages tend to be optimized either for broadband SED fitting or for image-plane morphology rather than providing a self-consistent treatment of both.  For rapid population studies and parameter space exploration, there thus remains a need for end-to-end tools that are fast to evaluate and easy to deploy while retaining physical interpretability.  And if the tool is to be widely used beyond a relatively small community of experts, then the reality of the current astronomical software ecosystem is that users seek software that is written in (or at least interfaces seamlessly with) Python \citep[e.g.,][]{Momcheva_2015,Astropy_2022}.

This paper aims to provide a tool that addresses the above considerations.  We present a semi-analytic jet model that is written in Python, designed to be computationally inexpensive, and which takes in physically interpretable input parameters and self-consistently produces both the internal jet structure as well as the corresponding synchrotron observables needed for SED and VLBI studies.  Our jet model is based on the prior semi-analytic work of \citet{Tchekhovskoy_2008} and \citet{Tsunetoe_2025}, and the corresponding codebase is hosted in a GitHub repository\footnote{\url{https://github.com/dpesce/ffjet}}, with the version used in this paper archived on Zenodo \citep{ffjet_zenodo}.

This paper is organized as follows.  \autoref{sec:Model} describes our semi-analytic jet model and some numerical implementation details.  In \autoref{sec:SED} we explore how the spectral energy distribution (SED) structure predicted by our jet model depends on the various input parameters.  \autoref{sec:Comparison} presents a comparison of the predictions from our jet model against observations of the \m87 and \sgra systems.  We summarize and conclude in \autoref{sec:Summary}.  Throughout this paper, we adopt a distance to \m87 of 16.8\,Mpc \citep{Blakeslee_2009,Bird_2010,Cantiello_2018,M87Paper6} and a distance to \sgra of 8.15\,kpc \citep{Reid_2019}.

\section{Model description} \label{sec:Model}

The model we use in this paper is effectively the same as that in \citet{Tsunetoe_2025}, which is based on the force-free jet construction initially developed by \citet{Tchekhovskoy_2008} and later refined by \citet[][see also \citealt{Broderick_2009}]{Gelles_2025}.  The force-free description is appropriate for the highly magnetized and low-density central ``funnel'' or ``spine'' of the jet, where the energy density in the electromagnetic field is much greater than the energy in the plasma.  GRMHD simulations consistently find that this region is well approximated by force-free electrodynamics \citep[e.g.,][]{McKinney_2006,Tchekhovskoy_2011}.

In this section, we provide an overview of the salient features of this model, but additional details are contained in the above papers.  We adopt a frame-dependent notation, in which primed quantities are measured in the fluid frame and unprimed quantities are measured in the lab frame.  Quantities that describe the electron population and emission are defined in the fluid frame and carry primes, while the electromagnetic field structure, bulk fluid motion, and observationally accessible intensities and frequencies are lab-frame quantities and thus unprimed.  Throughout the paper, we use $(r,\theta,\phi)$ to denote spherical coordinates and $(R,z,\phi)$ to denote cylindrical coordinates.

\subsection{Force-free jet}

We consider a stationary, axisymmetric, magnetically dominated (force-free) jet in the spacetime of a Kerr black hole with dimensionless spin parameter $a$ and mass $M$.  We express all quantities used to define the electromagnetic field structure in geometric units with $G=c=M=1$, such that the event horizon radius is given by

\begin{equation}
r_{\text{H}} = 1 + \sqrt{1-a^2} . \label{eqn:rH}
\end{equation}

\noindent We introduce physical lengths only when required for radiative transfer or energetics, using the gravitational radius

\begin{equation}
r_g = \frac{GM}{c^2} \approx \left( 1.477 \times 10^{5} \text{ cm} \right) \left( \frac{M}{M_{\odot}} \right) . \label{eqn:rg}
\end{equation}

\noindent We assume that the black hole is accreting at a rate $\dot{M}$, which we cast in terms of the scaled Eddington accretion rate,\footnote{In many studies, an effective radiative efficiency, $\eta_{\rm eff}=0.1$, is used for defining the Eddington mass accretion rate (e.g., \citealt{Narayan_Yi_1995,Yuan_Narayan_2014}). We do not include this factor.}

\begin{equation}
\dot{M} = \dot{m} \left( \frac{L_{\text{Edd}}}{c^2} \right) , \label{eqn:AccretionRate}
\end{equation}

\noindent where $L_{\text{Edd}} = 1.26 \times 10^{38} M$\,erg\,s$^{-1}$ is the Eddington luminosity. For computing images, we take the observer to be oriented at an inclination angle $i$ relative to the jet axis (which is taken to be aligned with the BH spin axis).  We use a simplified flat-ray approximation for carrying out radiative transfer, while retaining local relativistic effects (that enter through Kerr metric factors) to compute redshifts and frame conversions.

\subsubsection{Magnetic stream function and jet edge}

The jet shape is defined by the magnetic stream function, $\psi$, which is preserved along magnetic field lines.  Following \citet{Tchekhovskoy_2008}, we take the stream function in polar coordinates $(r,\theta)$ to be given by

\begin{equation}
\psi(r,\theta) \propto r^{2-2s} \big[ 1 - |\cos(\theta)| \big] , \label{eqn:StreamFunction}
\end{equation}

\noindent which provides an approximate solution (accurate to $\sim$10\%) to Maxwell's equations in the axisymmetric FF regime \citep[see also][]{Narayan_2007}.  The index $s$ sets the shape of the jet, such that the jet edge follows a power law with cylindrical radius $R \propto z^s$ at large radii.  We determine the proportionality constant in \autoref{eqn:StreamFunction} by fixing the total jet power $P_{\text{jet}}$, as described in \autoref{sec:SandJnorm} below.

In our implementation, the outer edge of the jet is defined by the magnetic field lines that pass through $\theta = \pi/2$ when $r = r_{\text{H}}$.  Because a magnetic field line necessarily follows a contour of constant magnetic stream function $\psi$, we can write an expression for the jet edge $r_{\text{jet}}^{\pm}(\theta)$ as

\begin{equation}
r_{\text{jet}}^{\pm}(\theta) = r_{\text{H}} \left( \frac{1}{1 \mp \cos(\theta)} \right)^{1/(2-2s)} , \label{eqn:JetProjRTheta}
\end{equation}

\noindent such that the jet region corresponds to $r \le r_{\text{jet}}^{+}(\theta)$ in the upper hemisphere and $r \le r_{\text{jet}}^{-}(\theta)$ in the lower hemisphere.

\subsubsection{Field line rotation rate}

The magnetic field lines rotate at an angular rate $\Omega$, which \citet{Gelles_2025} showed takes the form

\begin{equation}
\Omega(r,\theta) = \frac{a}{4} \left[ 1 + \sec^2\left( \frac{\theta_{\text{H}}(r,\theta)}{2} \right) \right]^{-1} \label{eqn:FieldLineOmega}
\end{equation}

\noindent for $0 < s < 1$.  Here, $\theta_{\text{H}}(r,\theta)$ is the polar angle at which the field line passing through $(r,\theta)$ intersects the horizon.  The field line rotation rate approaches the well-known value $\Omega \rightarrow a/8$ as $\theta_{\text{H}} \rightarrow 0$, which is half the horizon angular velocity $\Omega_{\text{H}} = a/(2 r_{\text{H}})$ to lowest order in $a$, as in the standard ``BZ monopole'' solution \citep{Blandford_1977}.

\begin{figure*}[t]
    \includegraphics[width=1.00\textwidth]{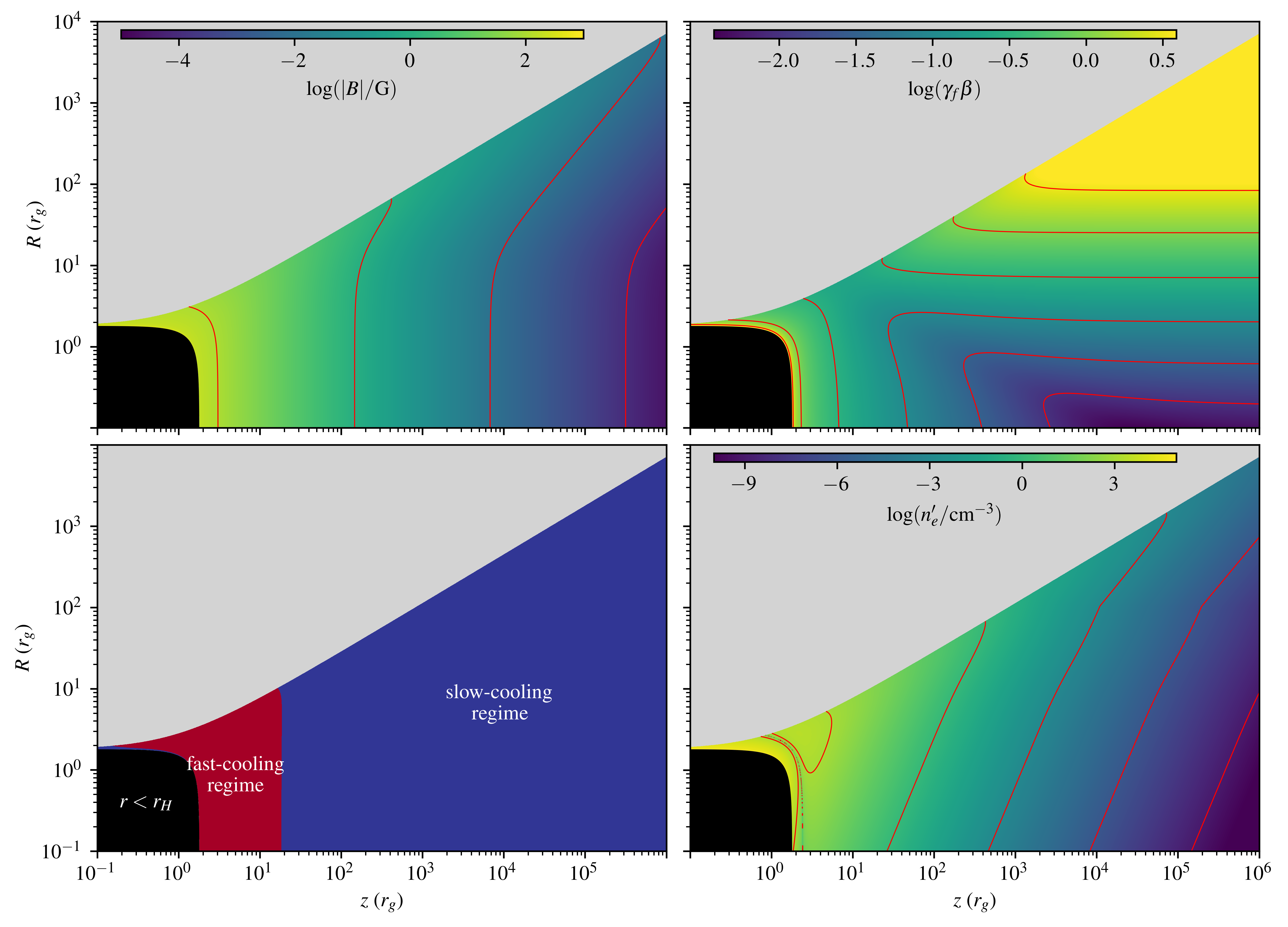}
    \caption{Cross-sections of the magnetic field strength (upper left), fluid velocity (upper right, plotted as $\gamma_f \beta = \sqrt{\gamma_f^2-1}$, which can take values between zero and infinity), electron number density (lower right), and electron cooling regimes (lower left) for our jet model, as a function of cylindrical coordinates $(R,z)$ and using the fiducial \m87-like parameters specified in \autoref{tab:Parameters}.  In each panel, the red contours correspond to the parameter values specified in the associated colorbar.  All panels share the same axis ranges, which are labeled in the lower-left panel.}\label{fig:map2d_4panel}
\end{figure*}

\subsubsection{Electromagnetic fields}

We determine the electromagnetic field following the prescription described in \citet{Gelles_2025}.  Defining the standard Kerr metric functions in Boyer-Lindquist coordinates,

\begin{subequations}
\begin{align}
\rho^2 & = r^2 + a^2 \cos^2(\theta) \\
\Delta & = r^2 - 2r + a^2 \\
\Sigma & = (r^2 + a^2)^2 - a^2 \Delta \sin^2(\theta) ,
\end{align}
\end{subequations}

\noindent we construct the lapse $\alpha$ as

\begin{equation}
\alpha = \sqrt{\frac{\rho^2 \Delta}{\Sigma}} \label{eqn:Lapse}
\end{equation}

\noindent and the metric determinant $\sqrt{-g}$ as

\begin{equation}
\sqrt{-g} = \rho^2 \sin(\theta) .
\end{equation}

\noindent The components of the magnetic field in the lab frame are then given by

\begin{subequations}
\begin{align}
B_r & = \frac{\alpha}{\sqrt{-g}} \sqrt{\frac{\rho^2}{\Delta}} \left( \pder{\psi}{\theta} \right) \\
B_{\theta} & = -\frac{\alpha}{\sqrt{-g}} \sqrt{\rho^2} \left( \pder{\psi}{r} \right) \\
B_{\phi} & = \alpha \sqrt{\frac{\Sigma \sin^2(\theta)}{\rho^2}} \left( \frac{I_{\text{pol}}}{2 \pi \Delta \sin^2(\theta)} \right) ,
\end{align}
\end{subequations}

\noindent where

\begin{equation}
I_{\text{pol}} = \mp 4 \pi \psi \Omega
\end{equation}

\noindent is the poloidal current (constructed such that the sign is negative in the upper hemisphere and positive in the lower hemisphere).  The magnetic field strength as a function of cylindrical coordinates, $R=r\sin(\theta)$, $z=r\cos(\theta)$, for an \m87-like jet is shown in the upper left panel of \autoref{fig:map2d_4panel}.

The electric field is set by enforcing ideal rotation of field lines relative to the local frame dragging.  Defining the frame-dragging frequency $\omega$,

\begin{equation}
\omega = \frac{2 a r}{\Sigma},
\end{equation}

\noindent the components of the electric field are given by

\begin{subequations}
\begin{align}
E_r & = - ( \Omega - \omega ) \frac{\alpha \Sigma \sin(\theta)}{\sqrt{\rho^2 \Delta} \sqrt{-g}} \left( \pder{\psi}{r} \right) \\
E_{\theta} & = - ( \Omega - \omega ) \frac{\alpha \Sigma \sin(\theta)}{\sqrt{\rho^2} \Delta \sqrt{-g}} \left( \pder{\psi}{\theta} \right) \\
E_{\phi} & = 0 .
\end{align}
\end{subequations}

\noindent The $E_{\phi} = 0$ condition is required for axisymmetry.

\subsubsection{Fluid velocity} \label{sec:FluidVelocity}

In a force-free setup, $\boldsymbol{E}$ and $\boldsymbol{B}$ are orthogonal, which implies that there exists a reference frame in which one of the fields vanishes.  It is conventional to choose the fluid frame to be the one in which $\boldsymbol{E}$ vanishes, the velocity for which must live in the plane spanned by $\boldsymbol{B}$ and $\boldsymbol{E} \times \boldsymbol{B}$.  We can thus express the velocity of the fluid as measured in the lab frame as

\begin{align}
\boldsymbol{v} & = \boldsymbol{v}_{\perp} + \boldsymbol{v}_{\parallel} \nonumber \\
& = v_{\perp} \left( \frac{\boldsymbol{E} \times \boldsymbol{B}}{|\boldsymbol{E} \times \boldsymbol{B}|} \right) + v_{\parallel} \left( \frac{\boldsymbol{B}}{|\boldsymbol{B}|} \right) .
\end{align}

\noindent The velocity component perpendicular to the magnetic field is the so-called ``drift velocity,'' and is given by

\begin{equation}
\boldsymbol{v}_{\perp} = \frac{\boldsymbol{E} \times \boldsymbol{B}}{B^2} .
\end{equation}

\noindent Force-free electrodynamics does not, by itself, fix the component of the plasma velocity parallel to $\boldsymbol{B}$.  However, \citet{Gelles_2025} showed that in the force-free limit a unique ``cold fluid'' velocity emerges once one imposes certain MHD regularity conditions.  We implement the parallel fluid velocity component using

\begin{equation}
v_{\parallel} = C_{\parallel} \sqrt{1 - v_{\perp}^2} .
\end{equation}

\noindent The coefficient $C_{\parallel}$ is field-line-dependent, and we compute it using the \citet{Gelles_2025} stagnation surface prescription.  Briefly: along each magnetic field line, we compute the radius of the stagnation surface (i.e., the radius at which the poloidal velocity vanishes), and we use a conserved energy integral (which is readily computable at the stagnation surface) to determine the value of $C_{\parallel}$ for the field line.  See \citet{Gelles_2025} for details.

Given the fluid velocity $v$ -- for which, when $c=1$, we have $\beta = v$ -- the fluid Lorentz factor $\gamma_f$ is

\begin{equation}
\gamma_f = \frac{1}{\sqrt{1 - v^2}}
\end{equation}

\noindent Following \citet{Tsunetoe_2025}, we apply two additional regularizations to the fluid velocity.  First, we suppress the $\gamma \beta$ product by a factor of 0.5, which is meant to emulate the effect of drag at the jet boundaries; \citet{Tsunetoe_2025} determined this value empirically by comparing the force-free velocity profile to GRMHD simulations.  Second, we impose a hard cap on the fluid Lorentz factor $\gamma_f \leq \gamma_{\infty}$ while preserving the fluid velocity direction, such that

\begin{equation}
\gamma_f \rightarrow \begin{cases}
\gamma_f , & \gamma_f < \gamma_{\infty} \\
\gamma_{\infty} , & \gamma_f \geq \gamma_{\infty} \\
\end{cases} .
\end{equation}

\noindent This cap serves to emulate fluid inertia, which is not otherwise captured by the force-free formalism.  The resulting fluid velocity structure for an \m87-like jet is shown in the upper right panel of \autoref{fig:map2d_4panel}.

\subsubsection{Poynting flux and jet power normalization} \label{sec:SandJnorm}

Given the magnetic field strength $B$ and the component of the fluid velocity $v_{\perp}$ that is perpendicular to the magnetic field, we determine the Poynting flux as

\begin{equation}
S = \frac{c}{4 \pi} B^2 v_{\perp} . \label{eqn:Poynting}
\end{equation}

\noindent We numerically integrate $S$ across both jet faces at large radii ($z = 10^8\,r_g$) to determine the jet power, and we rescale the magnetic field such that this jet power is fixed to be

\begin{equation}
P_{\text{jet}} = 1.4 a^2 \dot{M} c^2 , \label{eqn:JetPower}
\end{equation}

\noindent where the coefficient of 1.4 is appropriate for a prograde magnetically arrested accretion flow \citep{Tchekhovskoy_2011,Narayan_2022}.\footnote{The jet power is substantially less in the case of retrograde accretion. Furthermore, \cite{Cho_2025,Cho_2026} find that, in the case of radiatively inefficient accretion of gas from beyond the Bondi radius, there is no difference between prograde and retrograde rotation of the external gas, and the jet power $P_{\rm jet}$ lies in between the prograde and retrograde estimates in \cite{Tchekhovskoy_2011,Narayan_2022}.}  This jet power prescription also serves to determine the proportionality constant for the magnetic stream function (\autoref{eqn:StreamFunction}).

\subsection{Electron distribution function} \label{sec:eDF}

The force-free model describes the structure of the electromagnetic field and fluid within the jet, but the emission that we see ultimately comes from electrons whose properties must be prescribed separately.  \citet{Tsunetoe_2025} assumed that the electrons producing the observed synchrotron emission get their energy from the jet power, such that the energy density in emitting electrons $u_e'$ should be proportional to the Poynting flux $S$.  We retain this proportionality but modify the ansatz slightly, by assuming here that the electron energy density is given by

\begin{equation}
u_e' = \frac{h S}{c \gamma_f} , \label{eqn:ElectronEnergy}
\end{equation}

\noindent where $h \ll 1$ sets the fraction of the Poynting energy density that is converted into nonthermal energy in the emitting electrons per logarithmic interval in $r$; we emphasize that this ansatz describes continuous particle acceleration at every location within the jet, and that it implicitly assumes either the production of particles or an existing reservoir of non-emitting electrons available to be heated (i.e., electron number flux and energy flux are not conserved along the jet).  \autoref{eqn:ElectronEnergy} departs from the analogous expression in \citet{Tsunetoe_2025} by the addition of a factor of $\gamma_f$ into the denominator, which serves to express the energy density in the fluid frame, and which has the conceptual advantage that it is presumably the frame in which the electron energization takes place.

Following \citet{Tsunetoe_2025}, we assume that the nonthermal electrons are initially injected with a power-law distribution of Lorentz factors $\gamma'$,

\begin{equation}
f_{\text{inj}}(\gamma') \propto \begin{dcases}
0 , & \gamma' < \gamma_m' \\
\gamma'^{-p} , & \gamma_m' \leq \gamma' < \gamma_{\text{max}}' \\
0 , & \gamma_{\text{max}}' \leq \gamma' \\
\end{dcases} , \label{eqn:InjectedeDF}
\end{equation}

\noindent where $\gamma_m'$ and $\gamma_{\text{max}}'$ are the minimum and maximum Lorentz factors, respectively, and $p$ is the power-law index.  The normalization of \autoref{eqn:InjectedeDF} is such that $f_{\text{inj}}(\gamma')$ integrates to 1.  The number density of electrons $n_e'$ can then be determined by setting the integrated energy density equal to $u_e'$, which (assuming $p > 2$) yields

\begin{equation}
n_e' = \frac{u_e'}{m_e c^2} \frac{\big( p - 2 \big) \big( \gamma_{m}'^{1-p} - \gamma_{\text{max}}'^{1-p} \big)}{\big( p - 1 \big) \big( \gamma_{m}'^{2-p} - \gamma_{\text{max}}'^{2-p} \big)} . \label{eqn:ElectronNumberDensity}
\end{equation}

\noindent The electron number density for an \m87-like jet is shown in the lower right panel of \autoref{fig:map2d_4panel}.

We assume that electrons located at a distance $z$ down the jet have been cooling via synchrotron radiation for a characteristic timescale (in the fluid frame)

\begin{equation}
t_c' = \frac{z}{\gamma_f \beta_f c} , \label{eqn:CoolingTime}
\end{equation}

\noindent which we note again differs from the analogous expression in \citet{Tsunetoe_2025} by a factor of $\gamma_f$ (to allow for relativistic time dilation between the fluid and lab frames); there is also an extra factor of $\beta_f$ to replace $c$ in \citet{Tsunetoe_2025} by the more accurate fluid speed $v_f=\beta_f c$.  This cooling timescale is set by fluid advection and is the shortest of the relevant dynamical timescales along the jet.  The comoving expansion timescale $R/(dR/dt') = z/(s \gamma_f \beta c)$ exceeds $t_c'$ by a factor of $1/s \approx 1.7$, and diffusion of the emitting electrons across the jet is prevented by gyro radii that are ${\sim}10^{-5}$-$10^{-9}$ of the jet radius (so the electrons cannot escape out the sides of the jet before being advected downstream).

The cooling timescale is associated with a corresponding critical Lorentz factor,

\begin{equation}
\gamma_c' = \frac{6 \pi m_e c}{\sigma_T B'^2 t_c'} , \label{eqn:CoolingGamma}
\end{equation}

\noindent such that electrons with $\gamma' > \gamma_c'$ are able to cool to $\gamma' \approx \gamma_c'$ within the timescale $t_c'$.  Note that $B'$ here is the fluid-frame magnetic field strength (see \autoref{eqn:BFieldFluid} and \autoref{app:BfieldFluidFrame}).  The result of this cooling is a redistribution of the electron energies, with the new $f(\gamma')$ being well-approximated as a broken power-law,

\begin{equation}
f(\gamma') \propto \begin{dcases}
0 , & \gamma' < \gamma_1' \\
\gamma'^{-p_1} , & \gamma_1' \leq \gamma' < \gamma_2' \\
\gamma_2'^{p_2 - p_1} \gamma'^{-p_2} , & \gamma_2' \leq \gamma' < \gamma_3' \\
0 , & \gamma_3' \leq \gamma'
\end{dcases} . \label{eqn:eDF_gamma}
\end{equation}

\noindent Here, $\gamma_1'$ and $\gamma_3'$ are the minimum and maximum Lorentz factors, respectively, and $\gamma_2'$ is the Lorentz factor at which the ``break'' in the double power-law occurs; $p_1$ and $p_2$ are the two power-law indices.  It can be shown \citep[see, e.g.,][]{Sari_1998} that $p_2 = 1 + p$ and that $\gamma_3' = \gamma_{\text{max}}'$.  The values of $\gamma_1'$, $\gamma_2'$, and $p_1$ depend on whether the electrons are in a ``fast-cooling'' or ``slow-cooling'' regime \citep{Sari_1998}; see \autoref{app:ElectronCoolingRegimes} for details. Fast-cooling corresponds to $\gamma_c' < \gamma_m'$, and slow-cooling to $\gamma_c' > \gamma_m'$. The regions of the jet corresponding to these two cooling regimes for an \m87-like jet are shown in the lower left panel of \autoref{fig:map2d_4panel}.  We note that \autoref{eqn:eDF_gamma} describes a local, steady-state electron population that is implicitly assumed to be adiabatically cooling (though this cooling channel is not explicitly used to evolve the population); we discuss this assumption in more detail in \autoref{app:LocalEDF}.

As in \citet{Tsunetoe_2025}, we adopt an anisotropic electron distribution function, whereby electrons preferentially move parallel to magnetic field lines \citep[see also][]{Galishnikova_2023}.  The magnitude of this preference is controlled by the anisotropy parameter $\eta$, such that $\eta = 1$ corresponds to an isotropic distribution.  Given an angle $\xi'$ between an electron's motion and the magnetic field (measured in the fluid frame), we adopt the angular distribution function

\begin{equation}
f(\xi') \propto \big[ 1 + (\eta - 1) \cos^2(\xi') \big]^{-p_{\eta}/2} , \label{eqn:eDF_xi}
\end{equation}

\noindent where $p_{\eta}$ is a power-law index.  As in \citet{Tsunetoe_2025}, we make the approximation $\xi' = \theta_B'$ -- where $\theta_B'$ is the (fluid-frame) angle between the emitted synchrotron photons and the magnetic field -- which is appropriate for relativistic electrons.

We assume that the electron distribution function is separable in $\gamma'$ and $\xi'$, such that the joint distribution is given by

\begin{equation}
f(\gamma',\xi') = f(\gamma') f(\xi') , \label{eqn:eDF}
\end{equation}

\noindent and it is normalized such that

\begin{equation}
\frac{1}{2} \int_1^{\infty} d\gamma' \int_0^{\pi} f(\gamma', \xi') \sin(\xi') d\xi' = 1 .
\end{equation}

\noindent The distribution of electron number density is then just given by the product of $n_e'$ (\autoref{eqn:ElectronNumberDensity}) and $f(\gamma',\xi')$.  Note that the prescription used here differs somewhat from that in \citet{Tsunetoe_2025}, where there were two different values for $p_{\eta}$ for the two branches of \autoref{eqn:eDF_gamma}.  Here, we instead enforce that $p_{\eta}$ is the same in both branches, and we adopt a value of $p_{\eta} = 2$ throughout this paper.

\begin{deluxetable*}{lccc}
\tablecolumns{4}
\tablewidth{0pt}
\tablecaption{Jet model parameters\label{tab:Parameters}}
\tablehead{\colhead{Parameter} & \colhead{Description} & \colhead{\m87 value} & \colhead{\sgra value}}
\startdata
$M$ & black hole mass & $5.98 \times 10^9$\,M$_{\odot}$ & $4.07 \times 10^6$\,M$_{\odot}$ \\
$a$ & dimensionless spin parameter & $0.6$ & $0.86$ \\
$i$ & inclination angle & 163$^{\circ}$ & 26$^{\circ}$ \\
$\dot{M}$ & mass accretion rate & $8.91 \times 10^{-5}$\,$L_{\text{Edd}}/c^2$ & $8.57 \times 10^{-7}$\,$L_{\text{Edd}}/c^2$ \\
$s$ & jet collimation parameter & 0.6 & 0.6 \\
$p$ & injected eDF power-law index & 2.23 & 2.88 \\
$h$ & fraction of jet power that goes into heating nonthermal electrons & $1.14 \times 10^{-3}$ & $10^{-3}$ \\
$\eta$ & synchrotron anisotropy parameter & 0.03 & 0.01 \\
$p_{\eta}$ & power-law index for electron anisotropy component & 2 & 2 \\
$\gamma_{\infty}$ & maximum fluid Lorentz factor & 4 & 4 \\
$\gamma_m'$ & minimum Lorentz factor for injected power-law electrons & 10 & 10 \\
$\gamma_{\text{max}}'$ & maximum Lorentz factor for injected power-law electrons & $10^8$ & $10^8$ \\
\enddata
\tablecomments{Parameters describing the jet model used in this paper, along with fiducial values used for our \m87 and \sgra comparisons.  The fiducial parameter sets are determined from the SED-fitting analyses described in \autoref{app:SEDfitting} and \autoref{app:SEDfittingSgrA}.}
\end{deluxetable*}

\begin{figure}[t]
    \includegraphics[width=0.80\columnwidth]{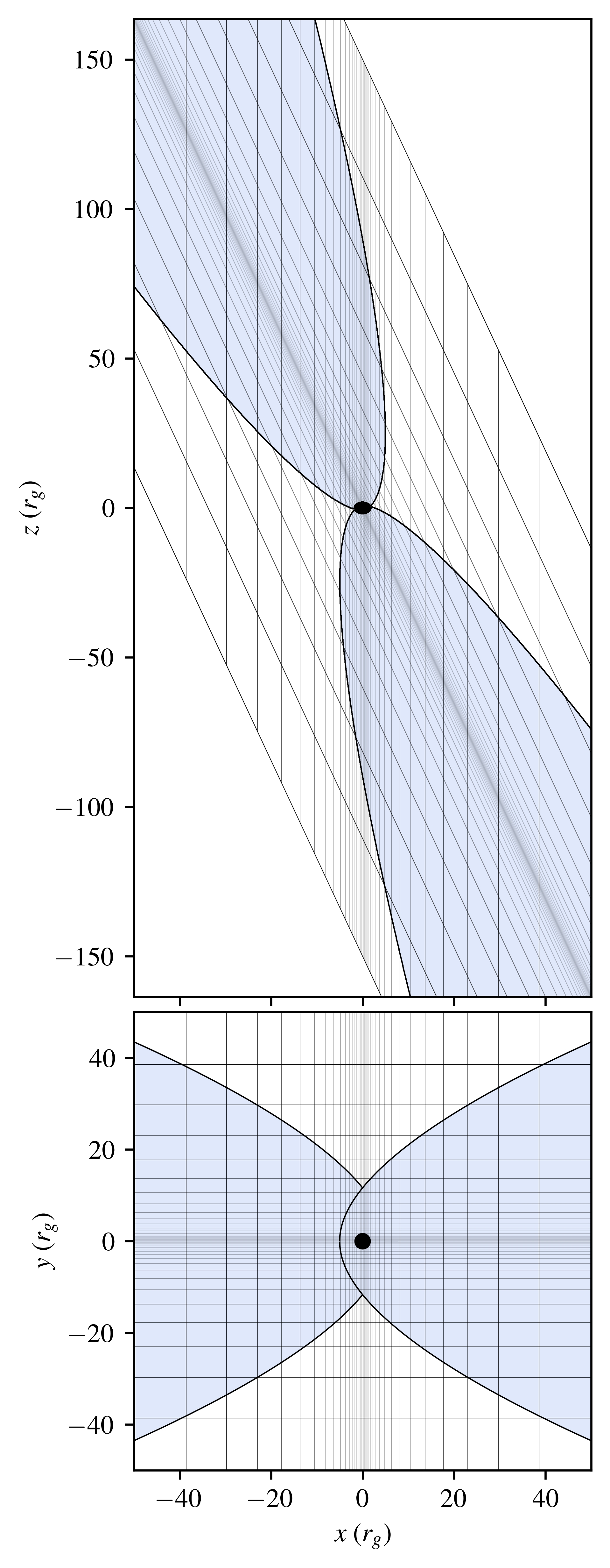}
    \caption{Example gridding scheme for jet image generation, with the grid cells traced out via the gray lines and the jet region highlighted in blue.  The grid is logarithmic in the observer's $x$, $y$, and $z$ coordinates -- such that the observer sees the image in the $(x,y)$-plane -- but is sheared to follow the jet axis.  The projected jet shape shown in the bottom panel is computed as described in \autoref{app:ProjectedJetShape}.}\label{fig:grid_2panel}
\end{figure}

\subsection{Radiative transfer} \label{sec:RT}

The equation for unpolarized relativistic radiative transfer is \citep[see, e.g.,][]{Dexter_2016}

\begin{equation}
\frac{d\mathcal{I}}{d\lambda} = \mathcal{J} - \mathcal{A} \mathcal{I} ,
\end{equation}

\noindent where $\mathcal{I} = I_{\nu} / \nu^3$ is the invariant intensity, $\mathcal{J} = j_I' / \nu'^2$ is the invariant emissivity, $\mathcal{A} = \nu' \alpha_I'$ is the invariant absorption coefficient, and $d\lambda = ds/\nu$ is a so-called affine parameter.  Our expressions for the synchrotron emissivity and absorption coefficients are described in \autoref{app:SynchEmisAbsorp}.

Switching to observer-frame quantities, we have

\begin{equation}
\frac{dI_{\nu}}{ds} = g^2 j_I' - g^{-1} \alpha_I' I_{\nu} , \label{eqn:RTEquation}
\end{equation}

\noindent where

\begin{equation}
g = \frac{\nu}{\nu'}
\end{equation}

\noindent is the redshift factor relating the frequency in the fluid frame to that in the observer frame (\autoref{eqn:RedshiftFactor}) and the coefficients $j_I'$ and $\alpha_I'$ are evaluated at the fluid-frame frequency $\nu' = \nu / g$.

Defining the optical depth

\begin{equation}
\tau \equiv \int \frac{\alpha_I'}{g} ds
\end{equation}

\noindent and source function

\begin{equation}
S_{\nu} \equiv g^3 \frac{j_I'}{\alpha_I'} ,
\end{equation}

\noindent the formal solution of the radiative transfer equation \citep[e.g.,][]{Rybicki_1979} then provides

\begin{equation}
I_{\nu}(\tau) = I_{\nu}(0) e^{-\tau} + \int_0^{\tau} e^{-(\tau - \tau')} S_{\nu}(\tau') d\tau' . \label{eqn:RTFormalSolution}
\end{equation}

Unlike \citet{Tsunetoe_2025} -- who conducted general relativistic radiative transfer in a Kerr spacetime -- in this paper we carry out radiative transfer along straight rays (i.e., not accounting for gravitational lensing by the black hole).  This simplification is made for both numerical speed and convenience, and because a flat-space approximation is reasonable along most of the length of the jet (i.e., outside $\sim$10\,$r_g$).

We implement our jet model numerically on a Cartesian grid in which the $z$-axis is aligned with the line-of-sight and the $(x,y)$-plane is aligned with the sky plane.  The grid spacing in all three dimensions is logarithmic, with the $(x,y)$-grid centered on the black hole and the $z$-grid centered on the jet axis, as illustrated in \autoref{fig:grid_2panel}.  Expressing the grid resolution as the number of points per dex (ppd) per grid dimension, we find that integrated flux densities are converged to ${\sim}0.3\%$ at 20\,ppd and to ${\lesssim}0.1\%$ at 40\,ppd (relative to computations carried out on a finer grid).

We numerically integrate \autoref{eqn:RTFormalSolution} across each cell $k$ by assuming that the source function is constant across the cell,

\begin{equation}
I_{k} = I_{k-1} e^{- \tau_{k}} + S_k (1 - e^{-\tau_k}) .
\end{equation}

\noindent To convert from intensity to flux density, we use

\begin{equation}
F_{\nu} = I_{\nu} \Omega ,
\end{equation}

\noindent where $\Omega$ is the solid angle subtended by an image pixel.

\section{Spectral energy distribution} \label{sec:SED}

\begin{figure*}[t]
    \includegraphics[width=0.90\textwidth]{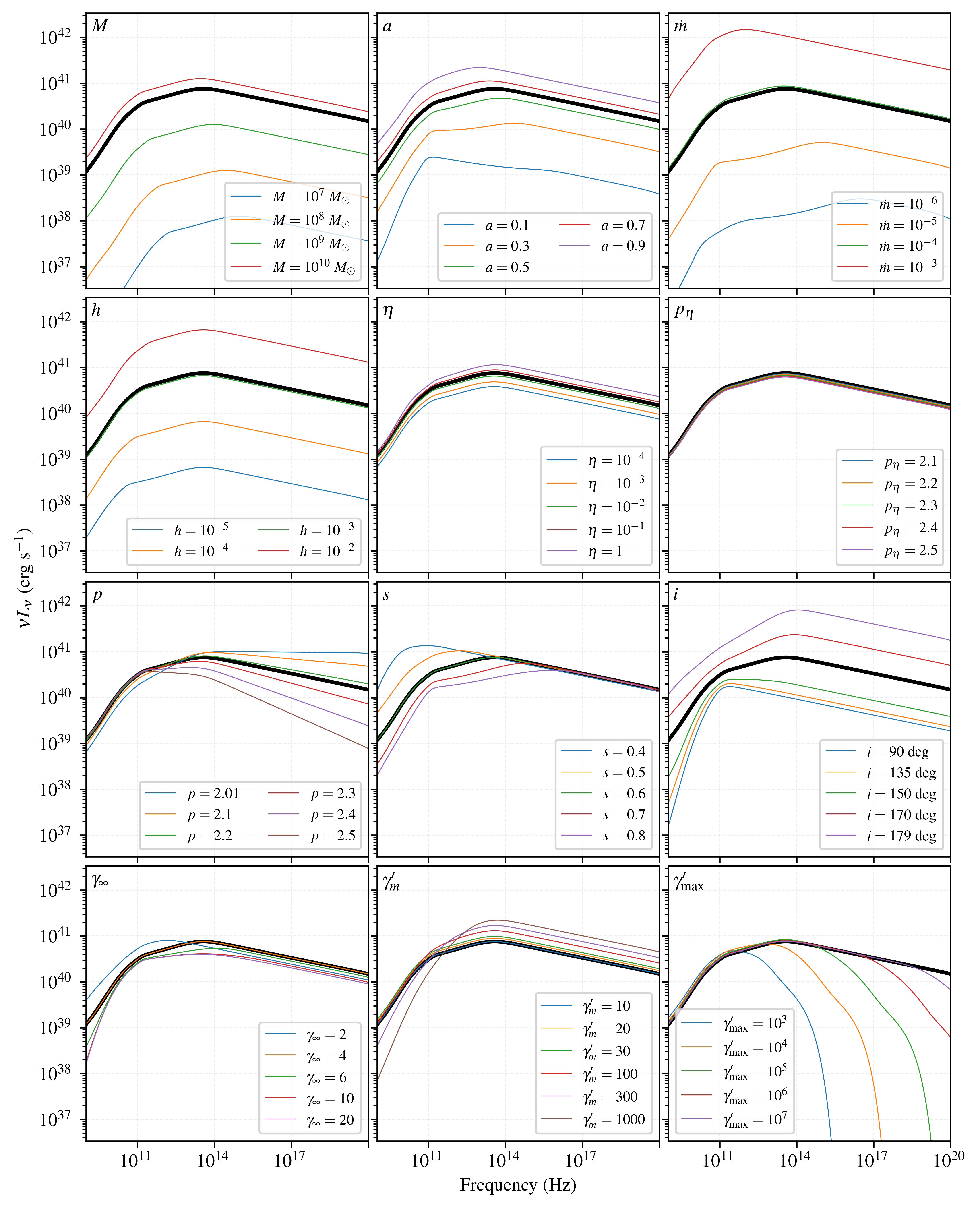}
    \caption{The impact of changing different model parameters on the SED, with each panel isolating the effect of a single parameter (labeled in the upper left-hand corner of the panel).  In each panel, the black curve shows the SED corresponding to our fiducial \m87 parameter set (\autoref{tab:Parameters}), and the colored curves show the SEDs corresponding to modifying the labeled parameter.  The SEDs are evaluated within $r \leq 10^5 r_g$.  All panels share the same axis ranges, which are labeled in the bottom left panel.}\label{fig:parameter_survey}
\end{figure*}

In this section we explore how the electromagnetic fields and emission physics in our jet model change as we vary model parameters, and in particular how these changes manifest in the observed spectral energy distribution (SED).  In addition to providing physical intuition, the approximate self-similarities derived in this section have a practical application in that they can replace explicit model evaluations when exploring large regions of parameter space for, e.g., population studies.  A gallery of the SEDs produced via surveying across each model parameter individually is shown in \autoref{fig:parameter_survey}, using the parameters for \m87 in \autoref{tab:Parameters} as our baseline model.  Unless otherwise noted, we hold the geometric parameters of the model -- i.e., the jet collimation parameter $s$ and inclination angle $i$ -- fixed for all scalings derived below.  We restrict our attention to $r < 10^5 r_g$ because the collimation profile changes beyond this radius in M87 \citep{Asada_2012} and in other sources \citep{Kovalev_2020,Burd_2022}.

\begin{figure*}[t]
    \includegraphics[width=1.00\textwidth]{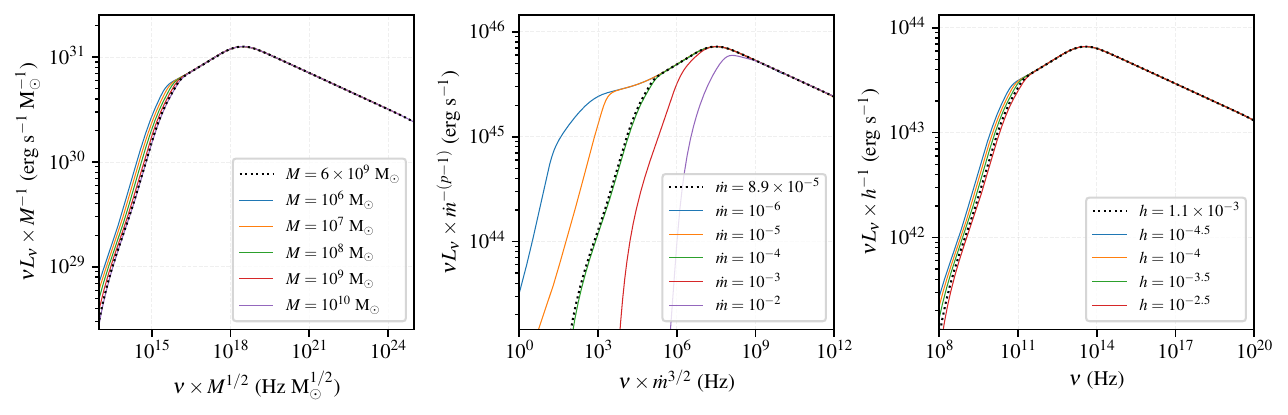}
    \caption{Illustrating the approximately self-similar nature of the jet model emission with black hole mass $M$ (left), accretion rate $\dot{m}$ (center), and jet heating fraction $h$ (right).  Each curve shows an SED computed using different values of $M$ (left panel, after rescaling per \autoref{eqn:SEDScalingMass}), $\dot{m}$ (center panel, after rescaling per \autoref{eqn:SEDScalingMdot}), or $h$ (right panel, after dividing $L_{\nu}$ by $h$).  In each panel, the SEDs are evaluated within $r \leq 10^5 r_g$.  Other than the parameter being varied, all model parameters take values corresponding to our fiducial \m87 parameter set (\autoref{tab:Parameters}); the SED corresponding to the fiducial value of the selected parameter is shown by the dashed black curve in each panel.}\label{fig:scale-free_SED}
\end{figure*}

\subsection{Physical scalings}

Because all spatial scales in the system scale with the black hole mass through $r_g \propto M$, and because the accretion rate scales with $M$ via the Eddington luminosity (\autoref{eqn:AccretionRate}), the total jet power (\autoref{eqn:JetPower}) will obey the proportionality

\begin{equation}
P_{\text{jet}} \propto a^2 \dot{M} \propto a^2 \dot{m} M . \label{eqn:ScalingJetPower}
\end{equation}

\noindent As described in \autoref{sec:SandJnorm}, the force-free electromagnetic field solution is rescaled such that the integrated Poynting flux equals $P_{\text{jet}}$.  Dimensionally, this rescaling implies that

\begin{equation}
B \propto P_{\text{jet}}^{1/2} r_g^{-1} \propto a \dot{m}^{1/2} M^{-1/2} ,
\end{equation}

\noindent up to geometric factors that depend on $a$ and the jet collimation index $s$.

Important electron Lorentz factors include the low-energy break of the injected distribution ($\gamma_m'$) and the critical cooling Lorentz factor ($\gamma_c'$), which itself obeys (\autoref{eqn:CoolingGamma})

\begin{equation}
\gamma_c' \propto B'^{-2} t_c'^{-1} \propto a^{-2} \dot{m}^{-1} . \label{eqn:ScalingGammaC}
\end{equation}

\noindent Characteristic synchrotron frequencies such as $\nu_c$ (\autoref{eqn:CharacteristicSynchrotronFreq}) scale linearly with the magnetic field and quadratically with the Lorentz factor of the emitting electrons \citep{Rybicki_1979},

\begin{equation}
\nu \propto B \gamma'^2 . \label{eqn:ScalingFreq}
\end{equation}

\noindent The frequency $\nu_m$ associated with $\gamma_m'$ will then scale as

\begin{equation}
\nu_m \propto B \gamma_m'^2 \propto \gamma_m'^2 a \dot{m}^{1/2} M^{-1/2} , \label{eqn:ScalingFreqNum}
\end{equation}

\noindent while the frequency $\nu_c$ associated with $\gamma_c'$ will scale as

\begin{equation}
\nu_c \propto B \gamma_c'^2 \propto a^{-3} \dot{m}^{-3/2} M^{-1/2} . \label{eqn:ScalingFreqNuc}
\end{equation}

\noindent The maximum injected Lorentz factor ($\gamma_{\text{max}}'$) also sets a high-frequency SED cutoff for synchrotron emission, but this parameter is estimated to be very large in M87* and other emission processes (e.g., inverse Compton emission) are likely to be important for describing the SED at these high frequencies (gamma-rays).

The electron density $n_e'$ scales directly with the jet heating fraction $h$ and with the Poynting flux, such that

\begin{equation}
n_e' \propto h a^2 \dot{m} M^{-1} . \label{eqn:ScalingElectronDensity}
\end{equation}

\noindent The synchrotron emissivity $j_{\nu}$ (\autoref{eqn:Emissivity}) depends primarily on the electron density and magnetic field strength,

\begin{equation}
j_{\nu} \propto n_e' B \propto h a^3 \dot{m}^{3/2} M^{-3/2} , \label{eqn:ScalingEmissivity}
\end{equation}

\noindent such that the emitted luminosity density in the optically thin regime follows

\begin{equation}
L_{\nu} \propto j_{\nu} r_g^3 \propto h a^3 \dot{m}^{3/2} M^{3/2} . \label{eqn:ScalingLuminosity}
\end{equation}

\noindent These scalings are sufficient to understand a number of structures seen in the SEDs produced by our jet model. We now consider the various panels in \autoref{fig:parameter_survey}.

\subsection{Self-similarity across black hole mass}

The critical cooling Lorentz factor $\gamma_c'$ does not scale with $M$ (\autoref{eqn:ScalingGammaC}), because the mass-dependence of $B^2$ and $t_c'^{-1}$ cancel out.  This independence of $\gamma_c'$ on $M$ means that both $\nu_m$ and $\nu_c$ share a common scaling with $M$ (see \autoref{eqn:ScalingFreqNum} and \autoref{eqn:ScalingFreqNuc}), which in turn permits the SED to be described as an approximately self-similar family of functions that is traversed by $M$.

Per \autoref{eqn:ScalingLuminosity}, the emitted luminosity scales with black hole mass as $L_{\nu} 
\propto M^{3/2}$.
Because all synchrotron frequencies share a common mass scaling of $M^{-1/2}$ (see \autoref{eqn:ScalingFreqNum} and \autoref{eqn:ScalingFreqNuc}), we can approximately convert an SED computed using some fiducial black hole mass $M_0$ into one with any other black hole mass via

\begin{equation}
F_{\nu}(\nu ; M) = \left[ \frac{M}{M_0} \right]^{3/2} F_{\nu}\left( \left[ \frac{M}{M_0} \right]^{1/2} \nu ; M_0 \right) , \label{eqn:SEDScalingMass}
\end{equation}

\noindent so long as we are in the optically thin regime (such that radiative transfer effects are not important).  The left panel of \autoref{fig:scale-free_SED} shows how this rescaling of the SED behaves across different choices of $M$.  We can see that the rescaling is nearly perfect at high frequencies, where the emission is optically thin.  It breaks down at lower frequencies, where the optical depth becomes important.

\subsection{Self-similarity across accretion rate}

The black hole spin parameter $a$ and accretion rate $\dot{m}$ are approximately degenerate through their impact on the total jet power (\autoref{eqn:ScalingJetPower}), which ensures that all downstream quantities are proportional to some power of the product $a^2 \dot{m}$.  However, because $a$ and $\dot{m}$ scale $\gamma_c'$ (\autoref{eqn:ScalingGammaC}) while leaving other important Lorentz factor thresholds (e.g., $\gamma_m'$) unchanged, these quantities impact how fast- and slow-cooled electrons are distributed within the jet.  The spin parameter is further complicated because it influences the geometry of the jet (through $r_{\text{H}}$; \autoref{eqn:JetProjRTheta}) and the fluid velocity structure (through $\Omega$; \autoref{eqn:FieldLineOmega}).  There is thus no clean scaling of the SED with $a^2 \dot{m}$ in the same way that there is with $M$.

However, because $\dot{m}$ impacts only the electron cooling structure and not the jet geometry, we can determine an approximate scaling of the SED with $\dot{m}$ (at fixed values of the other parameters) by restricting our focus to a particular cooling regime.  For example, let's consider the slow-cooling regime.  In this regime, the SED segment between $\nu_m$ and $\nu_c$ follows $F_{\nu} \propto \nu^{-(p-1)/2}$ (\autoref{eqn:ScalingSPN}), so we can write

\begin{equation}
F_{\nu} = F_{\nu_m} \left( \frac{\nu}{\nu_m} \right)^{-(p-1)/2} ,
\end{equation}

\noindent where $F_{\nu_m}$ is the value of $F_{\nu}$ at $\nu_m$.  We know that any fixed value of $F_{\nu}$ (such as $F_{\nu_m}$) will scale per \autoref{eqn:ScalingLuminosity} as $\dot{m}^{3/2}$, while $\nu_m$ itself will scale per \autoref{eqn:ScalingFreqNum} as $\dot{m}^{1/2}$.  So at fixed observed $\nu$, we have

\begin{equation}
F_{\nu} \propto \dot{m}^{(p+5)/4} \nu^{-(p-1)/2} ,
\end{equation}

\noindent which describes a vertical scaling of the SED at fixed $\nu$.  To track changes in the SED shape, we define a relative frequency variable $x \equiv \nu/\nu_c$, such that $\nu = x \nu_c$.  The scaling of $\nu_c$ with $\dot{m}$ is given by \autoref{eqn:ScalingFreqNuc} and goes as $\dot{m}^{-3/2}$.  Plugging this into the previous expression yields

\begin{equation}
F_{\nu} \propto \dot{m}^{(2p+1)/2} x^{-(p-1)/2} ,
\end{equation}

\noindent i.e., $F_{\nu}$ is approximately self-similar in $x$.  We can thus approximately convert an SED computed using some fiducial accretion rate $\dot{m}_0$ into one with any other accretion rate via

\begin{equation}
F_{\nu}(\nu ; \dot{m}) = \left[ \frac{\dot{m}}{\dot{m}_0} \right]^{(2p+1)/2} F_{\nu}\left( \left[ \frac{\dot{m}}{\dot{m}_0} \right]^{3/2} \nu ; \dot{m}_0 \right) . \label{eqn:SEDScalingMdot}
\end{equation}

\noindent A set of rescaled SEDs for different values of $\dot{m}$ is shown in the center panel of \autoref{fig:scale-free_SED}.  As the accretion rate decreases, the slow-cooling and optically thin regimes dominate a larger fraction of the jet, and this self-similar description becomes appropriate over an increasingly wide range of frequencies.

\subsection{Parameters that primarily rescale the SED}

The synchrotron emissivity is proportional to the electron density $n_e'$ and a function of the pitch-angle anisotropy factor $\eta$ (\autoref{eqn:Emissivity}).  Consequently, the SED scales directly with these parameters in the optically thin regime, without any modification to the frequency structure (because there is no $h$- or $\eta$-dependence in \autoref{eqn:ScalingFreq}).  This behavior can be seen in \autoref{fig:parameter_survey} as a simple vertical shift of the SED upon changing $h$ or $\eta$, at fixed values of the other parameters; meanwhile, the index $p_{\eta}$ has almost no effect.  The right panel of \autoref{fig:scale-free_SED} shows rescaled SEDs for different values of $h$, illustrating that this scaling is similar in quality to that with the black hole mass $M$.

\subsection{Parameters that reshape the SED}

The remaining parameters shown in \autoref{fig:parameter_survey} do not admit a clean rescaling in the same manner as $M$, $\dot{m}$, and $h$, but their effects on the SED structure can still be understood from how they modify the local emissivity (\autoref{eqn:Emissivity}), the redshift factor (\autoref{eqn:RedshiftFactor}), and the jet geometry (\autoref{eqn:StreamFunction}).

\subsubsection{Spectral index}

For a power-law distribution of injected electron Lorentz factors, the SED in the optically thin regime should follow standard behavior for synchrotron emission \citep[see, e.g.,][]{Sari_1998}, which for slow-cooling looks like

\begin{equation}
F_{\nu,\text{slow}} \propto
\begin{dcases}
\left( \frac{\nu}{\nu_m} \right)^{1/3} , & \nu < \nu_m \\
\left( \frac{\nu}{\nu_m} \right)^{-(p-1)/2} , & \nu_m \leq \nu \leq \nu_c \\
\left( \frac{\nu_c}{\nu_m} \right)^{-(p-1)/2} \left( \frac{\nu}{\nu_c} \right)^{-p/2} , & \nu_c < \nu
\end{dcases} , \label{eqn:ScalingSPN}
\end{equation}

\noindent and for fast-cooling looks like

\begin{equation}
F_{\nu,\text{fast}} \propto
\begin{dcases}
\left( \frac{\nu}{\nu_c} \right)^{1/3} , & \nu < \nu_c \\
\left( \frac{\nu}{\nu_c} \right)^{-1/2} , & \nu_c \leq \nu \leq \nu_m \\
\left( \frac{\nu_m}{\nu_c} \right)^{-1/2} \left( \frac{\nu}{\nu_m} \right)^{-p/2} , & \nu_m < \nu
\end{dcases} . \label{eqn:ScalingSPN_fast}
\end{equation}

\noindent In both cases, changing the value of $p$ modifies the spectral index of the SED, which can be seen in \autoref{fig:parameter_survey} and becomes especially evident at high frequencies.

\subsubsection{Collimation parameter}

The collimation parameter $s$ changes the jet geometry itself and therefore most strongly affects the low-frequency SED structure, which preferentially traces the bulk jet volume at large radii.

In the large-$r$ limit, magnetic field lines follow $R \propto z^s$ (\autoref{eqn:StreamFunction}) and the field itself is toroidally dominated, such that $B \sim B_{\phi} \propto R^{-1} \propto z^{-s}$.  The characteristic synchrotron cooling timescale will obey $t_c' \propto z$ (\autoref{eqn:CoolingTime}), because $\gamma_f = \gamma_{\infty}$ is constant at large radii.  We can then combine these scalings with those in \autoref{eqn:ScalingFreqNum} and \autoref{eqn:ScalingFreqNuc} to determine the behavior of the spectral breaks,

\begin{subequations}
\begin{align}
\nu_m & \propto z^{-s} \\
\nu_c & \propto z^{3s-2} .
\end{align} \label{eqn:FrequencyScalingWithS}
\end{subequations}

\noindent So $\nu_m$ always decreases towards larger radii, while $\nu_c$ only does so for $s < 2/3$.  For values of $s<1/2$, a large fraction of the jet will be in the fast-cooling regime, because $\nu_c$ falls much more rapidly with $z$ than $\nu_m$.

The change in jet shape with $s$ also impacts the size of the emitting region, and so we expect that it should affect the total amount of emission.  Decreasing $s$ makes the jet more strongly collimated, such that at fixed $z$ the cross-sectional area is smaller and the Poynting flux density is larger, enhancing the electron density and magnetic field strength and thus the synchrotron emissivity.  At fixed $z$ and jet power, $S \propto R^{-2}$, so that $n_e' \propto R^{-2}$ and $B \propto R^{-1}$, and in the optically thin part of the spectrum between $\nu_m$ and $\nu_c$ the emissivity at fixed frequency scales as $j_{\nu} \propto n_e' B^{(p+1)/2} \propto R^{-(p+5)/2}$ (see \autoref{app:LongitudinalProfile}).  The jet segment between $z$ and $2z$ has an emitting volume ${\sim}R^2 z$, so that it contributes

\begin{equation}
L_{\nu} \propto j_{\nu} R^2 z \propto z R^{-(p+1)/2} ,
\end{equation}

\noindent which increases as the jet becomes more collimated; this is the behavior seen in \autoref{fig:parameter_survey} at intermediate frequencies, where the SED rises as $s$ decreases.  In contrast, above the cooling break the electrons radiate essentially all of the injected energy within the time $t_c'$, so that the emitted power is set by the injected power (i.e., by $h P_{\text{jet}}$ and $p$) rather than by the geometry of the jet.  We can see this behavior in \autoref{fig:parameter_survey}, where the SED is largely unchanging with $s$ at high frequencies.

However, at frequencies for which synchrotron self-absorption is relevant, the emission is determined by the area of the surface where $\tau \sim 1$ (i.e., the photosphere).  The absorption coefficient is given by \autoref{eqn:SynchStokesIAbsorption}; for the power-law part of the distribution, it scales as

\begin{equation}
\alpha_I' \propto n_e' B^{(p+2)/2} \nu^{-(p+4)/2} \propto z^{-s(p+6)/2} \nu^{-(p+4)/2} .
\end{equation}

\noindent The optical depth then scales as $\tau \propto \alpha_I' R$, such that

\begin{equation}
\tau \propto \left( z^{s} \nu \right)^{-(p+4)/2} ,
\end{equation}

\noindent with the result that the $z$-location where $\tau \sim 1$ scales as

\begin{equation}
z_{\tau=1} \propto \nu^{-1/s} , \label{eqn:Tau1SurfaceZ}
\end{equation}

\noindent and thus that the synchrotron self-absorption turnover frequency $\nu_{\text{sa}}$ scales as

\begin{equation}
\nu_{\text{sa}} \propto z^{-s}, \label{eqn:SelfAbsorptionScalingWithS}
\end{equation}

\noindent just like $\nu_m$.

Interpreting the scalings given by \autoref{eqn:FrequencyScalingWithS} and \autoref{eqn:SelfAbsorptionScalingWithS} in light of the SEDs in \autoref{fig:parameter_survey} requires considering the radial structure of the jet, because these scalings only describe local emission while the SEDs are integrated out to radii of $10^5 r_g$.  The SEDs in \autoref{fig:parameter_survey} exhibit one spectral break for $s \lesssim 0.5$ and two spectral breaks for $s \gtrsim 0.5$.  The $s=0.5$ value (corresponding to a parabolic jet) is a special one, because this value is where both $\nu_m$ and $\nu_c$ share the same scaling of $z^{-0.5}$.  For $s < 0.5$, $\nu_c$ decreases more steeply with $z$ than $\nu_m$ or $\nu_{\text{sa}}$ does, leading to a rapid transition from the low-frequency self-absorbed regime to the high-frequency optically thin regime that manifests as just a single apparent turnover frequency in the integrated SED.

For $s > 0.5$, the SEDs in \autoref{fig:parameter_survey} exhibit two spectral breaks.  The lower-frequency break corresponds to where the innermost radii have transitioned into the self-absorbed regime\footnote{There is an additional spectral break at frequencies even below those plotted in \autoref{fig:parameter_survey}, corresponding to the point at which the outermost radii (and thus all radii contributing to the plotted SED) have also transitioned into the self-absorbed regime}.  The higher-frequency spectral break corresponds to $\nu_c$, which changes only modestly with $z$ for these values of $s$ (thus manifesting as a relatively sharp spectral break) and which increases with $s$ (per \autoref{eqn:FrequencyScalingWithS}).

\subsubsection{Inclination angle and terminal Lorentz factor}

Neither the inclination angle $i$ nor the terminal fluid Lorentz factor $\gamma_{\infty}$ modifies the intrinsic electromagnetic structure of the jet, but both quantities influence the SED through their impact on the redshift factor (\autoref{eqn:RedshiftFactor}).   We can write the Doppler part of the redshift factor $\delta$ in terms of the angle $\theta$ between the local fluid velocity and the line-of-sight,

\begin{equation}
\delta \equiv \frac{g}{\alpha} = \frac{1}{\gamma_f \big(1 - \beta \cos(\theta) \big)} \approx \frac{2 \gamma_f}{1 + \theta^2 \gamma_f^2} , \label{eqn:DopplerRedshift}
\end{equation}

\noindent where in the final equality we have taken the relativistic ($\gamma_f \gg 1$) and small-angle ($\theta \ll 1$) limits.  \autoref{eqn:DopplerRedshift} implies that  emission will be strongly boosted (i.e., $\delta \gg 1$) in regions of the jet where $\theta \lesssim 1/\gamma_f$, and that for a fixed $\theta$ the Doppler factor is maximized near $\gamma_f \approx 1/\theta$.  The most obvious impact of this Doppler boosting is that the approaching jet is brighter than the receding jet, a well-known effect that is also evident in \autoref{fig:M87_observations} and which serves to generically increase the total amount of emission seen in the SEDs in \autoref{fig:parameter_survey} as $i$ increases from $90$ to $180$ degrees (i.e., as the approaching half of the jet becomes more aligned with the line-of-sight).

The terminal fluid Lorentz factor $\gamma_{\infty}$ sets the maximum fluid velocity, and it therefore affects quantities that depend on $\gamma_f$.  Increasing $\gamma_{\infty}$ narrows the beaming cone ($\theta \propto 1/\gamma_f$) and thus increases the corresponding Doppler boosting, but it simultaneously decreases the intrinsic emissivity because $j_{\nu} \propto u_e' \propto \gamma_f^{-1}$ (\autoref{eqn:ElectronEnergy}).  Because $\gamma_c' \propto B'^{-2} t_c'^{-1} \propto \gamma_f^3$ (\autoref{eqn:BFieldFluid}, \autoref{eqn:ScalingGammaC}) scales with $\gamma_{\infty}$, decreasing $\gamma_{\infty}$ also moves the cooling break to lower frequencies and therefore shifts power away from the highest-frequency part of the SED.  At inclinations for which the jet points nearly along the line of sight, the Doppler beaming and emissivity-altering effects compete with one another, such that increasing $\gamma_{\infty}$ can either enhance or suppress the observed emission depending on which parts of the jet remain within the Doppler cone.

\m87's inclination of $i=163$\,degrees corresponds to an angle of $\theta = 17 \text{\,degrees} \approx 0.3$\,radians between the jet axis and the line of sight, such that the critical Lorentz factor is $\gamma_{\text{crit}} \approx 1 / \theta \approx 3.4$.  For $\gamma_{\infty} > \gamma_{\text{crit}}$, the beaming cone is already sufficiently narrow that additional concentration through increasing $\gamma_{\infty}$ does not increase the observed emission.  However, the intrinsic emissivity does still continue to decrease with increasing $\gamma_{\infty}$, and the cooling break moves to higher frequencies.  Both of these effects are evident in \autoref{fig:parameter_survey}, which exhibits a monotonically decreasing SED with increasing $\gamma_{\infty}$ for $\gamma_{\infty} \geq 4$, as well as a high-frequency spectral break that moves to higher frequencies for increasing $\gamma_{\infty}$.

For $\gamma_{\infty} < \gamma_{\text{crit}}$, the inner jet regions (which contribute to the high-frequency SED) become under-boosted relative to the outer jet regions, and the intrinsic emissivity increase is not enough to compensate for the weaker beaming and the downward shift of the cooling break.  We thus see a more pronounced shift of the SED in this case, with the increased low-frequency emission coming at the expense of decreased high-frequency emission.

\subsubsection{Electron distribution function Lorentz factor limits}

The parameters $\gamma_m'$ and $\gamma_{\text{max}}'$ set the minimum and maximum Lorentz factors, respectively, for the injected nonthermal electron population (\autoref{eqn:InjectedeDF}).  Changing these parameters thus modifies the fraction of the nonthermal electrons that have lower or higher energies, which necessarily also reweights the SED accordingly.  This effect can be seen in \autoref{fig:parameter_survey}, whereby changing $\gamma_m'$ or $\gamma_{\text{max}}'$ in a way that increases the high-frequency emission also causes a reduction in the low-frequency emission, and vice versa.

For the minimum injected Lorentz factor $\gamma_m'$, \autoref{eqn:ScalingFreqNum} gives $\nu_m \propto \gamma_m'^2$. Correspondingly, we see that the location of the low-frequency break moves rapidly as a function of $\gamma_m'$. The peak of the SED is not located at this break, but rather at the cooling break. \autoref{eqn:ElectronNumberDensity} gives $n_e' \propto \gamma_m'^{-1}$ (assuming $\gamma_{\text{max}}' \gg \gamma_m'$ and $p > 2$), such that $F_{\nu} \propto n_e' \propto \gamma_m'^{-1}$.  The consequence for the SED depends on the cooling regime.  In the slow-cooling regime, the SED peaks at $\nu_c$ and we have $F_{\nu_m} \propto n_e'$, so we can use \autoref{eqn:ScalingSPN} to determine the scaling for the SED peak to be

\begin{equation}
(\nu L_{\nu})_{\text{peak}} \propto \nu_c F_{\nu_c} \propto \nu_c F_{\nu_m} \left( \frac{\nu_c}{\nu_m} \right)^{-(p-1)/2} \propto \gamma_m'^{p-2} . \label{eqn:PeakSEDScalingGammaM}
\end{equation}

\noindent For our fiducial value of  $p = 2.23$, \autoref{eqn:PeakSEDScalingGammaM} gives $(\nu L_{\nu})_{\text{peak}} \propto \gamma_m'^{0.23}$.  In the fast-cooling regime, the SED peaks at $\nu_m$  and we have $F_{\nu_c} \propto n_e'$, so we use \autoref{eqn:ScalingSPN_fast} to determine the scaling for the SED peak,

\begin{equation}
(\nu L_{\nu})_{\text{peak}} \propto \nu_m F_{\nu_m} \propto \nu_m F_{\nu_c} \left( \frac{\nu_m}{\nu_c} \right)^{-1/2} \propto \gamma_m'^0 , \label{eqn:PeakSEDScalingGammaM_fast}
\end{equation}

\noindent i.e., the SED peak is almost entirely independent of $\gamma_m'$.  In general, the SED peak will be determined by some combination of \autoref{eqn:PeakSEDScalingGammaM} and \autoref{eqn:PeakSEDScalingGammaM_fast}, both of which vary only weakly with $\gamma_m'$.  This weak dependence is evident in \autoref{fig:parameter_survey}, where the SED peak changes by only a factor of $\sim$3 over two orders of magnitude in $\gamma_m'$.

The $\gamma_{\text{max}}'$ parameter controls the upper energy cutoff of the injected eDF, with a corresponding typical synchrotron emission frequency that scales as (see \autoref{eqn:ScalingFreq})

\begin{equation}
\nu_{\text{max}} \propto B \gamma_{\text{max}}'^2 .
\end{equation}

\noindent So as $\gamma_{\text{max}}'$ decreases, the high-frequency cutoff of the SED occurs at lower frequencies, which is evident in \autoref{fig:parameter_survey}.

\section{Comparison to data} \label{sec:Comparison}

In this section we illustrate how the jet model can be used to make predictions for real sources, specifically \m87 and \sgra.  Because extensive multi-wavelength SED measurements are available for both \m87 and \sgra, we carry out full multi-parameter inference to determine the best-fitting set of fiducial parameters for both sources, which we list in \autoref{tab:Parameters}.  We compare the resulting broadband SEDs and some representative radio images to existing observations.

\subsection{M87 jet comparison}

Our fiducial \m87 parameter set is determined primarily by the SED fit described in \autoref{app:SEDfitting}, which provides the joint posterior distribution over eleven model parameters given the 45 SED measurements compiled in \citet{M87_MWL}.  As specified in \autoref{tab:FittedParameters}, we assign informative priors only for the black hole mass and accretion rate, both of which are based on EHT measurements \citep{M87Paper6,M87Paper7}.  Several of the parameters (e.g., $\eta$, $\gamma_{\infty}$, $\gamma_m'$, $\gamma_{\text{max}}'$) are only weakly constrained, while others exhibit strong degeneracies (e.g., between black hole spin $a$ and the electron heating fraction $h$), such that no single set of per-parameter best-fit median values is itself a good fit to the data.  We instead construct the fiducial set by choosing round values for those parameters to which the SED is only weakly sensitive, and then re-fitting the sharply constrained parameters with the weak-parameter choices held fixed.

We further inform the weakly-SED-constrained parameters using the observed \m87 jet morphology (\autoref{fig:M87_observations}).  VLBI images at 8 and 15\,GHz from \citet{Nikonov_2023} show that the jet remains bright out to projected distances of ${\gtrsim}10^4\,r_g$, which within our model requires the fluid speed to approach a modest asymptotic Lorentz factor $\gamma_{\infty} \lesssim 5$ by $z \approx 10^3\,r_g$.  We thus retain $\gamma_{\infty} = 4$ for our fiducial model, which is the same value used by \citet{Tsunetoe_2025}.  We set the black hole spin to $a = 0.6$ -- which lives along the $a$-$h$ posterior degeneracy -- and we retain the anisotropy parameter $\eta = 0.03$ and maximum injected Lorentz factor $\gamma_{\text{max}}' = 10^8$ values from \citet{Tsunetoe_2025}.  We set the minimum injected Lorentz factor to $\gamma_m' = 10$ (which is consistent with the posterior and which has negligible impact on the predicted images), and we fix the jet collimation rate to $s = 0.6$, which matches the value measured by \citet{Asada_2012}.  We use $p = 2.23$ for the fiducial injected power-law index value -- corresponding to the best-fit value given the other fiducial parameter choices -- which is relatively tightly constrained by the X-ray SED observations.  Despite imposing an uninformative (isotropic) prior on the inclination angle, the SED-informed posterior distribution strongly constrains this parameter to $i = 159.7^{+3.7}_{-4.4}$ degrees, in good agreement with prior VLBI constraints \citep{M87Paper5,Mertens_2016,Walker_2018}; we adopt an inclination angle of $i = 163^{\circ}$ for our fiducial parameter value.

Given the above choices, we re-fit the three parameters to which the SED is most sensitive -- i.e., the black hole mass, the accretion rate, and the heating fraction $h$ -- obtaining best-fit values of $M = 5.98 \times 10^9$\,\msun, $\dot{m} = 8.91 \times 10^{-5}$, and $h = 1.14 \times 10^{-3}$, respectively.  \autoref{tab:Parameters} lists the full set of fiducial \m87 parameters.

\subsubsection{Representative M87 radio images}

\begin{figure*}[t]
    \includegraphics[width=1.00\textwidth]{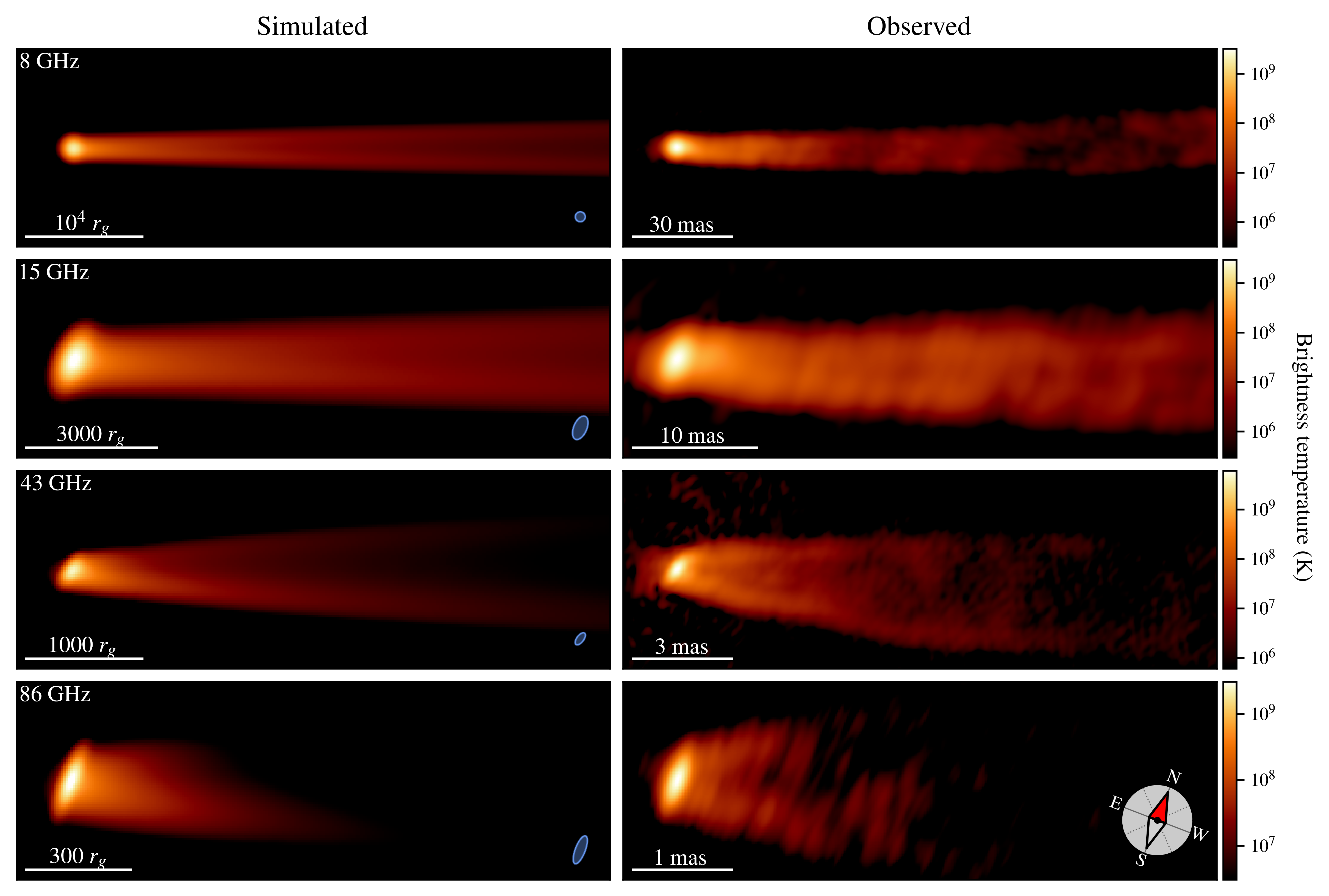}
    \caption{This figure has been reproduced with some modifications from \citet{Tsunetoe_2025}, and shows comparisons between modeled (left) and observed (right) images of the jet in \m87.  The left column shows simulated images at frequencies of (from top to bottom) 8, 15, 43, and 86\,GHz, while the right column shows the images from corresponding VLBI observations of \m87.  The 8 and 15\,GHz images come from \citet{Nikonov_2023}, the 43\,GHz image comes from \citet{Walker_2018}, and the 86\,GHz image comes from \citet[][using observations originally reported in \citealt{Lu_2023}]{Kim_2025}.  Each simulated image has been convolved with an elliptical Gaussian whose FWHM is shown in the lower right corner; the parameters of these convolving Gaussians match those of the corresponding restoring beams used for the VLBI images in the right column.  Each image also includes emission from an ADAF component -- the same model shown in \autoref{fig:SED_comparison_M87}, evaluated at the fiducial $M$ and $\dot{m}$ values -- which contributes flux densities of $\sim$0.03, $\sim$0.06, $\sim$0.17, and $\sim$0.35\,Jy at 8, 15, 43, and 86\,GHz, respectively.  Because the ADAF emission at these frequencies originates from the innermost few tens of gravitational radii, we render it in the displayed images as an unresolved core component.  All images have been rotated so that the \m87 jet points approximately to the right of the image, which corresponds to the compass rose orientation shown in the lower right corner of the figure.}\label{fig:M87_observations}
\end{figure*}

Because our model incorporates a number of modifications with respect to that presented in \citet{Tsunetoe_2025}, we first look to confirm that the model can still adequately produce images of the \m87 jet that comport with observations.  In \autoref{fig:M87_observations} we reproduce Fig.\,4 from \citet{Tsunetoe_2025} using the fiducial parameter set and model from the present paper.  This figure compares the predicted model images against VLBI observations of the \m87 jet at 8\,GHz, 15\,GHz, 43\,GHz, and 86\,GHz.  We continue to see reasonable agreement between the model predictions and the observations, even given the modifications to the model.

More quantitatively, the longitudinal jet intensity profile exhibits the scaling $I_{\nu} \propto z^{-s(p+3)/2} \approx z^{-1.57}$ for \m87 (see \autoref{app:LongitudinalProfile}).  This power-law index lives within the range of ${\sim}-1.5$ to ${\sim}-1.7$ from \citet[][see their Fig.~6]{Walker_2018} using observations at 43\,GHz on scales of several milliarcseconds (i.e., a few hundred to a few thousand $r_g$). It is also consistent with the broader range of ${\sim}-1$ to ${\sim}-3.5$ measured by \citet{Burd_2022} across multiple different blazars.

\subsubsection{Broadband M87 SED}

\begin{figure}[t]
    \includegraphics[width=1.00\columnwidth]{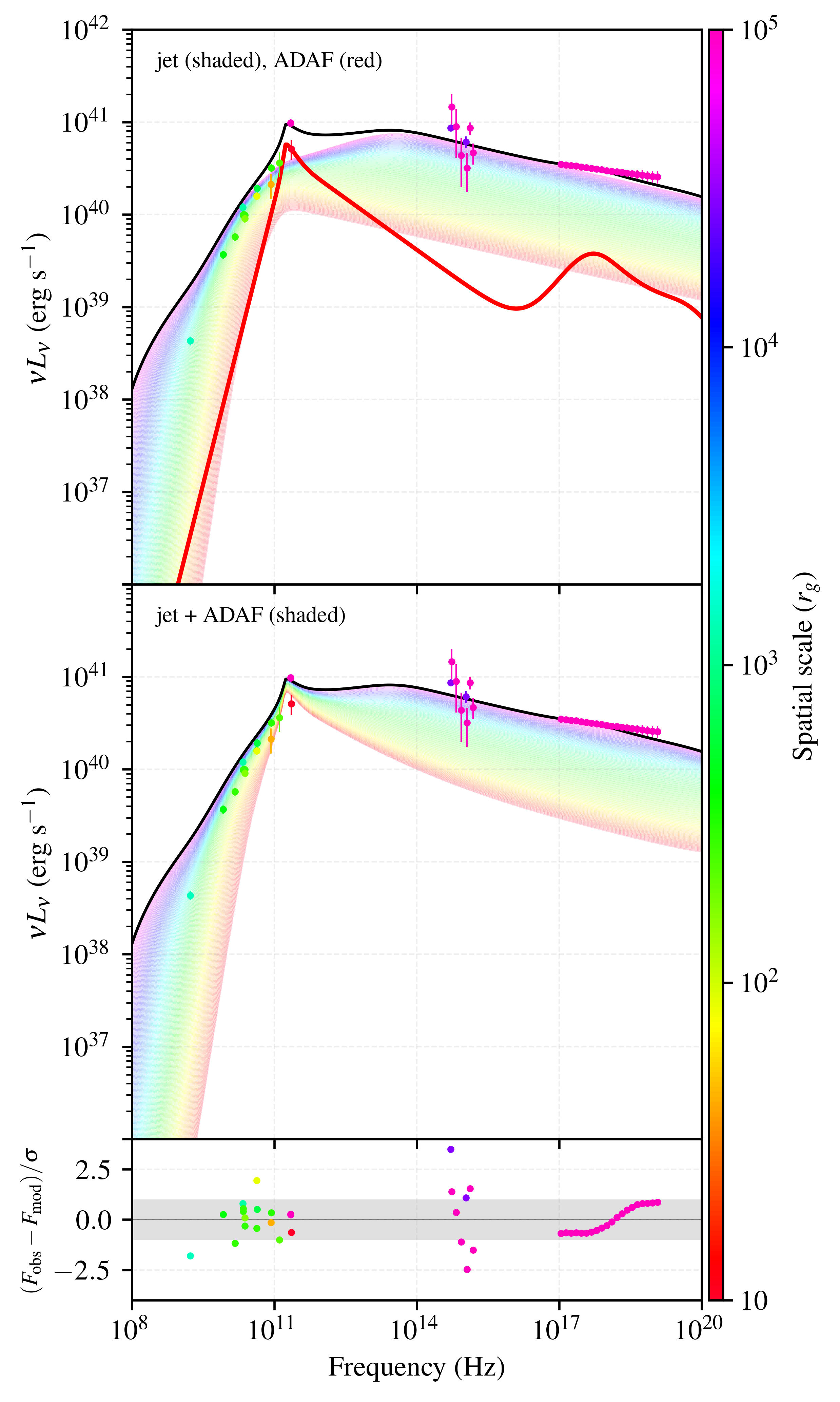}
    \caption{Model SED for \m87.  In the top panel, the rainbow-shaded region shows the predicted SED from our jet model over a range of spatial scales from $r \leq 10 r_g$ (in red) out to $r \leq 10^5 r_g$ (in magenta).  The red curve shows the predicted SED from the ADAF model described in \citet{Pesce_2021} -- using the same black hole mass and accretion rate as we use for our jet model (see \autoref{tab:Parameters}) -- and the black curve shows the sum of the ADAF and jet SEDs.  The middle panel is identical to the top panel, except that the red ADAF curve has been removed and the rainbow-shaded region now instead shows the combined jet and ADAF emission within each spatial scale.  The overlaid points \text{in both upper panels} represent real measurements, which are also colored according to the spatial scale they probe; these measurements come from \citet{M87_MWL}.  The bottom panel shows the normalized residuals $(F_{\text{obs}} - F_{\text{mod}})/\sigma$ for the fiducial model, with the gray band marking the $\pm 1 \sigma$ region.}\label{fig:SED_comparison_M87}
\end{figure}

We next compare our model to the multi-wavelength compilation of \citet{M87_MWL}, who assembled measurements of \m87's SED spanning radio through high-energy gamma rays (though our comparison only spans through the X-ray, as our model does not include the inverse Compton emission that is likely responsible for the high-energy SED structure).  The SED model and measurements for \m87 are shown in \autoref{fig:SED_comparison_M87}; we convert the measurements to $L_{\nu} = 4 \pi D^2 F_{\nu}$ using our adopted distance of $D = 16.8$\,Mpc.

The flux density measurements represented by the data points in \autoref{fig:SED_comparison_M87} are made on a variety of spatial scales, associated with the angular resolutions available to the facilities observing at different wavelengths; these different spatial scales are indicated by the colors of the data points.  The measured SED is thus not only a function of frequency, but also a function of the spatial scale out to which the measurements probe.  We capture this spatial scale dependence in the model SED by integrating the jet emission at each frequency out to a range of radii from $r_{\text{max}} = 10\,r_g$ (in red) out to $r_{\text{max}} = 10^5\,r_g$ (in magenta), represented by the colored band in \autoref{fig:SED_comparison_M87}.  The model naturally produces a frequency-dependent emitting region, with lower-frequency emission predominantly originating at large radii (and thus exhibiting a strong dependence on $r_{\text{max}}$) while higher-frequency emission is primarily produced in the inner jet region.

Because our jet model does not include emission from the accretion flow -- which is expected to contribute significantly to the flux density of \m87 at millimeter wavelengths -- we also incorporate the SED corresponding to the advection-dominated accretion flow (ADAF) model from \citet[][itself derived primarily from \citealt{Mahadevan_1997} and \citealt{Narayan_Yi_1995}]{Pesce_2021}.  We initialize the ADAF model using the same black hole mass and accretion rate as we use for our jet model (see \autoref{tab:Parameters}), and the corresponding SED is shown as the red curve in \autoref{fig:SED_comparison_M87}.  The sum of the ADAF and $r_{\text{max}} = 10^5\,r_g$ jet SEDs is shown as the black curve in \autoref{fig:SED_comparison_M87}.  We see that the ADAF does indeed contribute to the total SED primarily at millimeter wavelengths, becoming subdominant to the jet emission at both lower and higher frequencies.

Comparing the model to the measurements, we see that the combined jet+ADAF SED for \m87 reproduces well the flux density observed across the full sampled frequency range and associates that flux density with the correct spatial scales. A particularly interesting case is the behavior around a wavelength of 1\,mm ($\nu \approx 2 \times 10^{11}$\,Hz), where the model predicts that EHT observations on event-horizon scales (red data point) should be dominated by the ADAF emission while ALMA observations on much larger spatial scales (magenta data point) should contain comparable amounts of ADAF and jet emission.  This approximately equal split between horizon-scale and extended emission comports with the structure reported by the EHT Collaboration \citep{M87Paper4}.

The residual panel in \autoref{fig:SED_comparison_M87} quantifies the degree to which the data and model agree; our fiducial model achieves a value of $\chi^2 = 46.2$ across 45 measurements.  The largest remaining residuals are associated with the optical SED measurements near $\nu \approx 10^{15}$\,Hz, with the most significant single deviation being a ${\sim}35\%$ (${\sim}3.5\sigma$) underprediction of the HST optical flux density.  The nearby optical measurements scatter about the model with alternating signs at the $1$-$2.5\sigma$ level, with adjacent measurements typically disagreeing with one another by more than their quoted measurement uncertainties.  There is also some clear (though ${\lesssim}1\sigma$) systematic residual structure across the X-ray band, reflecting a small difference in the local SED slope between the model and data.  Both of these differences could plausibly arise from source variability -- which violates the steady-state assumption of our model\footnote{We note that the apparent transit time from $r\sim r_g$ to $r\sim 10^5r_g$ in M87 is a decade, even after allowing for superluminal motion, so any intrinsic variability in the source will manifest as additional unmodeled residuals when attempting to fit spectral data from such vastly different scales using a single time-steady model.} -- or the difficulty of separating the nuclear emission from that of the host galaxy at these wavelengths.

\subsection{Sgr A* jet estimates}\label{Sec:SgrA*}

Unlike in \m87, the presence -- let alone the morphology -- of any jet that may be originating from \sgra is substantially less certain.  The existence of a jet launched from \sgra has been proposed based on SED considerations, particularly as an explanation for the excess low-frequency radio emission observed alongside a ``submillimeter bump'' \citep[e.g.,][]{Serabyn_1997,Falcke_1998,Falcke_Markoff_2000,Yuan_2002,Yuan_2003}, but no unambiguous image of such a jet has ever been produced.  The comparison we present in this section should therefore be interpreted more cautiously than the \m87 case, with our goal being to demonstrate how our model can be used to assess whether a jet in \sgra is compatible with the observed SED, and to predict what such a jet might look like.

Similar to our approach for \m87, our fiducial parameter set for \sgra is derived from a fit to the quiescent-state SED, as described in \autoref{app:SEDfittingSgrA}.  The fit holds several weakly-constrained parameters fixed at values comparable to those used for \m87 -- i.e., $\eta = 0.01$, $p_{\eta} = 2$, $\gamma_{\infty} = 4$, $\gamma_m' = 10$, and $\gamma_{\text{max}}' = 10^8$ -- and we adopt a black hole mass prior of $M = (4.14 \pm 0.08) \times 10^6$\,M$_{\odot}$ \citep{Do_2019,GRAVITY_2019}.  To enforce a jet-like solution, we also fix $h = 10^{-3}$ (see \autoref{app:SEDfittingSgrA} for details).  We otherwise permit the remaining parameters to vary under uninformative priors.  After posterior exploration, we select the round value $s = 0.6$ (i.e., matching the \m87 value) and re-optimize the remaining parameters at that value to determine the fiducial parameter set, which is listed in \autoref{tab:Parameters}.

Our fiducial accretion rate of $\dot{m} \approx 9 \times 10^{-7}$ falls within the $\dot{m} \approx 5$--$9 \times 10^{-7}$ range from \citet{SgrA_Paper5}, and our fiducial inclination of $i \approx 30$\,degrees is similarly consistent with the constraints presented by the EHT \citep{SgrA_Paper5,SgrA_Paper8} and GRAVITY \citep{GRAVITY_2018} collaborations.  The fiducial value for the injected power-law index $p \approx 2.9$ is steeper than for \m87, and this steep index is known to be required to explain the infrared and X-ray SED behavior \citep[see, e.g.,][]{Ozel_2000,Yuan_2003,Broderick_2009b,Kimura_2022,vonFellenberg_2025}.\footnote{We note that requiring different values of $p$ for \m87 and \sgra is not physically unreasonable.  A comprehensive analysis of power-law indices observed in gamma-ray burst jets -- including both prompt emission (which is the most direct analog to our jet model) and afterglows, as well as blazars (a reasonable analogue) and pulsar wind nebulae -- indicates a wide distribution of $p$ in all these objects \citep{Shen_2006,Kumar_2015}.}  The fiducial spin value of $a = 0.86$ is set by the specific choice of $s$, but we emphasize that this parameter is otherwise unconstrained; the spin posterior for \sgra demonstrates that it is effectively unconstrained, with substantial posterior probability across the full prior range.

\begin{figure}[t]
    \includegraphics[width=1.00\columnwidth]{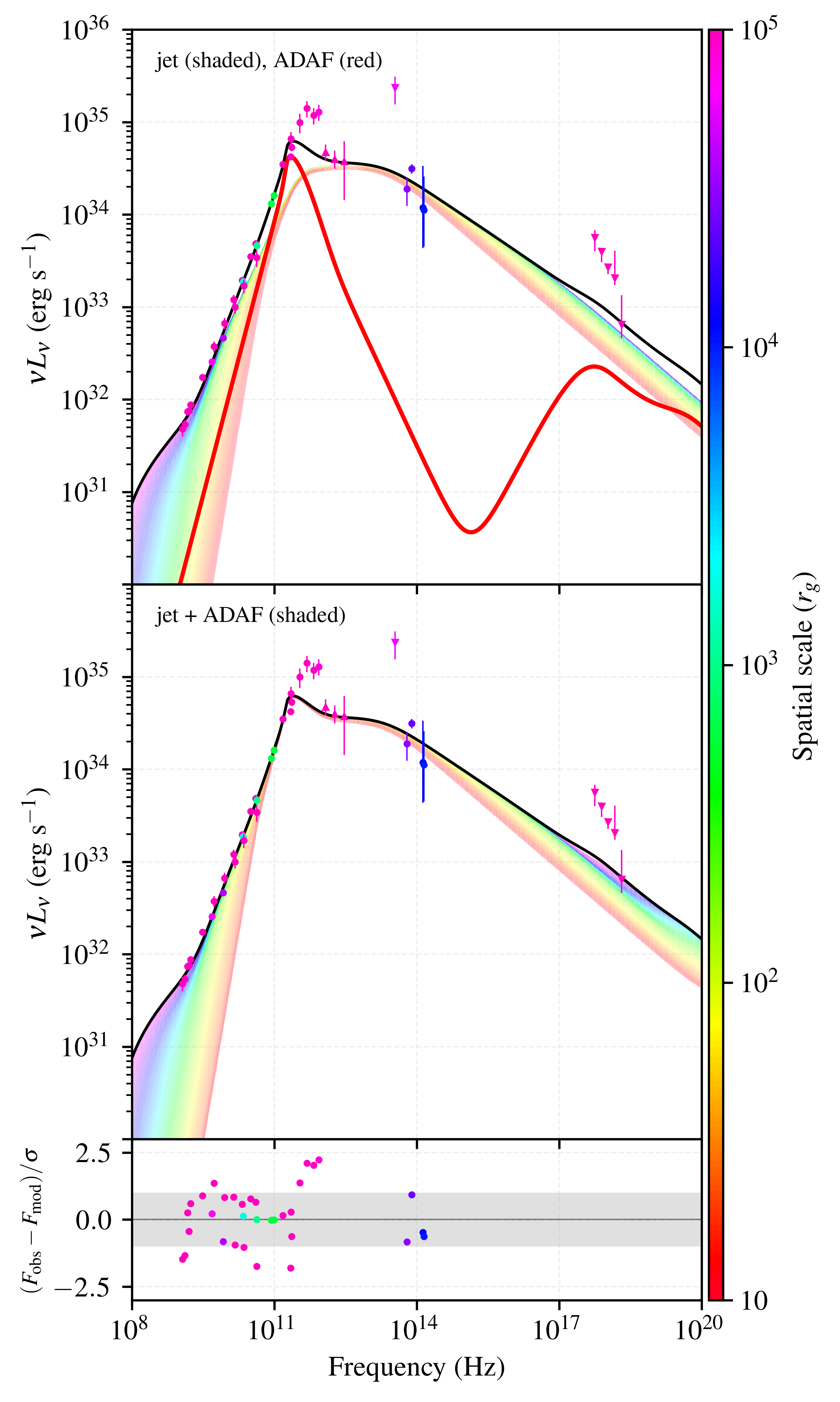}
    \caption{Same as \autoref{fig:SED_comparison_M87}, but showing here the SED for \sgra.  The overlaid measurements come from the papers listed in \autoref{tab:SEDMeasurementsSgrA}; triangle symbols indicate limits and are not included in the bottom residual panel.}\label{fig:SED_comparison_SgrA}
\end{figure}

\subsubsection{Broadband Sgr A* SED}

The primary observational evidence for a potential jet in \sgra comes from its SED, which we show in \autoref{fig:SED_comparison_SgrA}.  As in the \m87 comparison, the rainbow-shaded band shows the jet emission from our model integrated out to a range of spatial scales, the red curve shows the corresponding ADAF contribution, and the black curve shows the sum of the two.  The measurements shown  in \autoref{fig:SED_comparison_SgrA} come from a heterogeneous set of facilities and angular resolution scales, many of which have been compiled in \citet[][see \autoref{app:SEDMeasurements}]{SgrA_Paper2}; we have converted the measurements to $L_{\nu} = 4 \pi D^2 F_{\nu}$ using our adopted distance of $D = 8.15$\,kpc \citep{Reid_2019}.

Unlike for \m87 -- where low-frequency VLBI observations probe spatial scales close to the black hole -- the interstellar scattering along the line of sight to \sgra imposes an angular resolution limit that worsens as the square of the wavelength, restricting low-frequency observations to large spatial scales (e.g., \citealt{Johnson_2018} and references therein).  Constraints on the SED within $r_{\text{max}} \lesssim 2 \times 10^3 r_g$ are thus only available at millimeter wavelengths from VLBI observations with the EAVN, GMVA, and EHT \citep{SgrA_Paper2}.  Nevertheless, \autoref{fig:SED_comparison_SgrA} shows that our fiducial model provides a quantitatively acceptable match to the observed \sgra flux density at radio through (sub)millimeter wavelengths, achieving $\chi^2 = 37$ across 33 measurements and with no limit violated by more than 0.6$\sigma$.  The largest residuals are ${\sim}2\sigma$ underpredictions of the submillimeter flux, which may arise from the intrinsic variability of \sgra (the discrepant data points are primarily ALMA measurements made by \citealt{Bower_2019} within a short time period of one another).  In our fiducial model SED, the outer jet accounts for the rising centimeter-wave spectrum, while the inner jet and ADAF together produce the spectral steepening and broad peak around the millimeter/submillimeter ``bump.''  In this sense, the model is consistent with the long-standing interpretation that the millimeter excess in \sgra is produced by a compact synchrotron component associated with the accretion flow -- the ADAF -- very close to the black hole \citep{Narayan_1995}, supplemented with jet emission which progressively dominates at lower frequencies \citep{Falcke_1998}.

The high-frequency SED for \sgra is complicated by its large intrinsic variability, with changes in the flux density reaching factors of several hundred in the X-ray \citep[e.g.,][]{Baganoff_2001,Neilsen_2013}, and the origin of this high-frequency emission is still debated.  The quiescent-state X-ray emission seems to be well-described by thermal bremsstrahlung originating from plasma on spatial scales extending out to the Bondi radius \citep[${\sim}10^6$\,$r_g$; see, e.g.,][]{Quataert_2002,Baganoff_2003}, while the flaring state emission originates on more compact spatial scales and seems to be better-described by synchrotron and/or Compton processes \citep[e.g.,][]{Dodds-Eden_2009,Ponti_2017,Scepi_2022}.  Variability of the quiescent-state X-ray emission indicates that $\lesssim$10\% of it originates on compact spatial scales \citep{Neilsen_2013,Neilsen_2015}.

Our jet model assumes steady-state and contains only synchrotron emission, so it is particularly unsuited for describing the high-frequency \sgra SED.  Nevertheless, attempting to match the infrared observations from \citet{GRAVITY_2020} in the context of our jet model, while also retaining consistency with the radio observations, requires parameter choices that produce a steep decrease in flux density with $\nu$ at high frequencies.  In our fiducial model, this steepening is provided by the power-law index $p = 2.88$ (compare with \citealt{vonFellenberg_2025}), together with the low heating fraction ($h = 10^{-3}$) and minimum injected Lorentz factor ($\gamma_m' = 10$).  The X-ray portion of the SED in \autoref{fig:SED_comparison_SgrA} is thus dominated by synchrotron emission from the jet, with bremsstrahlung from the ADAF contributing at the ${\sim}20\%$ level.  This level of jet contribution exceeds the ${\lesssim}10\%$ allowed by the variability constraints discussed above, though we note that lowering the maximum injected Lorentz factor to $\gamma_{\text{max}}' \lesssim 3 \times 10^4$ would reduce the jet contribution without changing the fit to the rest of the SED.

\begin{figure*}[t]
    \includegraphics[width=1.00\textwidth]{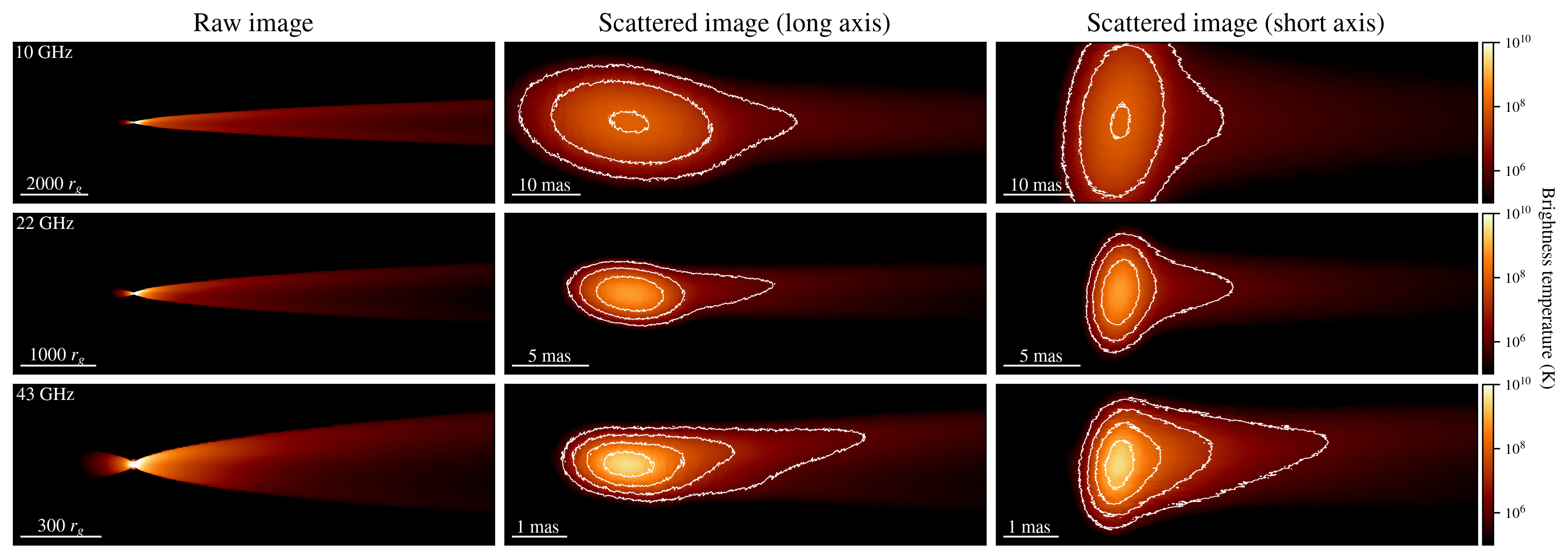}
    \caption{Model images for a potential jet in \sgra at 10\,GHz (top row), 22\,GHz (middle row), and 43\,GHz (bottom row), computed using the fiducial parameter set listed in \autoref{tab:Parameters}.  The left column shows the raw image predicted by the model, while the right two columns show the images after additionally applying the effects of interstellar scattering \citep[using the model from][]{Johnson_2018} along the line of sight to the Galactic Center.  The center column shows scattered images assuming that the jet is aligned primarily along the long axis of the scattering kernel, and the right column shows scattered images assuming   that the jet is aligned primarily along the short axis of the scattering kernel.  In the right two columns, the white contours are plotted at brightness temperature values of $10^6$, $10^7$, $10^8$, and $10^9$\,K.}\label{fig:SgrA_images}
\end{figure*}

\subsubsection{Radio images}

\autoref{fig:SgrA_images} shows predicted images of the hypothetical \sgra jet at frequencies of 10, 22, and 43\,GHz.  Because the emission from \sgra is known to be heavily affected by interstellar scattering at these frequencies \citep[e.g.,][]{Davies_1976,Bower_2006}, we also show versions of the images that have been processed using the scattering model from \citet{Johnson_2018}.  Given the unknown orientation of the jet with respect to the scattering kernel, we adopt two different relative orientations in which the jet points approximately parallel to (middle panels) and approximately perpendicular to (right panels) the major axis of the scattering kernel.

We see that at these radio frequencies, the broad scattering kernel dominates the brightest image structure, with the jet only becoming evident at brightness levels that are a factor ${\gtrsim}10^2$ fainter than the peak image brightness.  This brightness level is comparable to that of the refractive substructure observed at these frequencies \citep{Gwinn_2014}, and it appears to be fainter than any structure reconstructed in existing VLBI observations of \sgra \citep[e.g.,][]{Johnson_2018,Issaoun_2019}.  Previous simulations of \sgra have concluded that next-generation radio facilities (such as the ngVLA and ngEHT) will likely be necessary to detect its jet emission in image reconstructions \citep{Chavez_2024}.

Despite a lack of image reconstructions, the size of the \sgra emission structure at radio frequencies has nevertheless been constrained by VLBI observations.  \citet{Johnson_2018} reported the intrinsic structure of \sgra at these frequencies to have a full width at half maximum (FWHM) size that follows approximately

\begin{equation}
\theta_{\text{FWHM}} = 40 \text{\,$\mu$as} \left( \frac{\lambda}{1 \text{\,mm}} \right) , \label{eqn:SgrA_FWHM}
\end{equation}

\noindent with no evidence for significant anisotropy across a frequency range from $\sim$10\,GHz to $\sim$200\,GHz \citep[see also][]{Issaoun_2021}.

In \autoref{fig:FWHM_vs_frequency}, we compare \autoref{eqn:SgrA_FWHM} to measurements from our fiducial \sgra model; we have taken the model FWHM to be twice the radius out to which the emission must be integrated to accumulate half of the total flux seen within $10^5$\,$r_g$.  We find that our model FWHM roughly follows the $1/\nu$ scaling predicted by \autoref{eqn:SgrA_FWHM} up to frequencies of $\sim$200\,GHz (above which the size of the image structure becomes comparable to the size of the black hole shadow feature).  Below $\sim$200\,GHz, our predicted FWHM is systematically smaller than the measured value; this disagreement could simply reflect a modest level of model inaccuracy, or it could potentially indicate that emission from the accretion flow also contributes to the observed size of \sgra \citep[as interpreted by, e.g.,][]{Johnson_2018}.

\subsubsection{Discussion}

The absence of any clear evidence for a jet in radio images of \sgra has been a longstanding mystery.  While the predictions of our model (\autoref{fig:SgrA_images}) confirm that scatter-broadening could hide the jet, it appears that slightly deeper observations or more careful analysis of existing observations might reveal the underlying jet, especially at 43 GHz (and 86 GHz; see \citealt{Issaoun_2019, Issaoun_2021}).  In particular, the model predicts that non-Gaussian image structure should become evident at 43\,GHz once a brightness temperature sensitivity of ${\sim}10^7$\,K is achieved, which should be readily accessible to the ngVLA \citep{Selina_2024}.

\begin{figure}[t]
    \includegraphics[width=1.00\columnwidth]{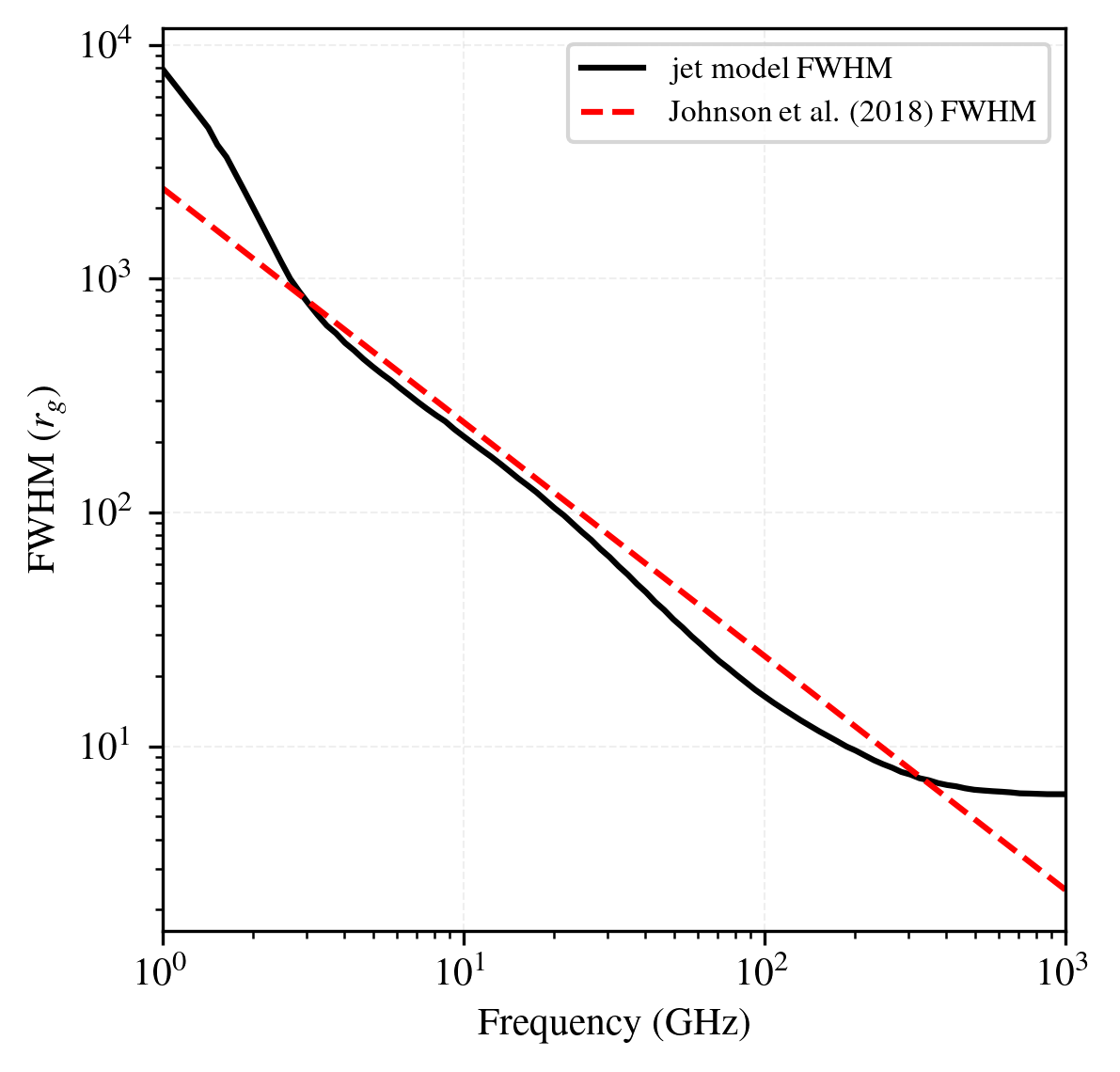}
    \caption{Comparison of the FWHM size of the emission structure in our fiducial \sgra jet model (black curve) to that measured by \citet{Johnson_2018} using VLBI observations of \sgra from $\sim$10\,GHz to $\sim$200\,GHz (red dashed line).}\label{fig:FWHM_vs_frequency}
\end{figure}

Could it be that \sgra has an unusually weak jet?  Since jet power is proportional to $a^2$ (\autoref{eqn:JetPower}), a slowly-spinning black hole would produce a correspondingly weak jet.  In our \sgra SED fit, the spin parameter is poorly constrained (see \autoref{app:SEDfittingSgrA}), and the high spin value in our fiducial parameter set is a consequence of our adopted $s=0.6$ collimation parameter value rather than being a data requirement.  Alternatively, perhaps the viewing angle $i$ is such that the observed flux is highly Doppler de-beamed.  With any such explanation, we will need to find an alternate explanation for the radio excess in \sgra's SED, which is clearly inconsistent with a thermal ADAF model (see the red curve in \autoref{fig:SED_comparison_SgrA}).  \citet{Ozel_2000} have shown that, by invoking a population of nonthermal electrons in the ADAF, both the radio spectrum of \sgra and the intrinsic angular size of the source can be simultaneously explained (compare their Figures 4 and 5 with \autoref{fig:SED_comparison_SgrA} and \autoref{fig:FWHM_vs_frequency} here).

\section{Summary and conclusions} \label{sec:Summary}

In this paper we have presented a model for the structure of and emission from astrophysical jets.  Up to a small number of modest modifications, our model is the same as that described in \citet{Tsunetoe_2025}, which itself is based on the force-free jet construction initially developed by \citet{Tchekhovskoy_2008} and later refined by \citet{Gelles_2025}; we retain the nonthermal electron heating prescription from \citet{Tsunetoe_2025}, but with small modifications to make the model more relativistically self-consistent.  Our primary contribution is in the exploration of this model's parameter space -- particularly as it impacts the predicted jet SED structure -- and in the comparison of the model's predictions against observations of the \m87 and \sgra systems.  We also provide an open-source Python implementation\footnote{\url{https://github.com/dpesce/ffjet}} \citep{ffjet_zenodo} that can be used to produce model images and SEDs.  The exact code runtime is system-dependent, but a single-frequency image takes about $\sim$1 second to run on a single CPU core of a 2021-era Macbook Pro, while a broadband SED evaluation takes several seconds using 8 CPU cores on the same machine.  These runtimes are low enough that posterior exploration is feasible on timescales of hours using just a personal computer.

In \autoref{sec:SED} we have explored how the SED structure predicted by our jet model depends on the various input parameters.  We find that the SED is approximately self-similar in the black hole mass parameter, such that modifying the black hole mass (given fixed values for the other model parameters) is nearly equivalent to shifting and scaling the SED.  A similar degree of self-similarity is present for the parameter $h$ (which describes the fraction of the jet power that goes into heating the nonthermal electron population) and, to a lesser extent, the mass accretion rate onto the black hole.  All of these self-similar relations become more exact in regions of parameter space where the emission is optically thin (e.g., at high frequencies and at low accretion rates).

In \autoref{sec:Comparison} we have compared the predictions from our jet model against the observed radio images and broadband SEDs of the \m87 and \sgra systems.  For \m87, we have carried out a full posterior exploration of the model parameter space, fitting our model to the broadband SED (\autoref{app:SEDfitting}) and demonstrating that it can quantitatively reproduce the observed SED (i.e., best-fit $\chi^2_{\nu} \approx 1$) across frequencies from radio through X-ray and on spatial scales from horizon-scale out to ${\sim}10^5$ gravitational radii.  Our fiducial parameter set simultaneously reproduces the VLBI images of the \m87 jet spanning ${\sim}$one order of magnitude in frequency and ${\sim}$two orders of magnitude in spatial scale, as well as the observed longitudinal intensity profile.

We have also compared the predictions of our jet model to observations of \sgra, for which the presence of a jet remains debated.  We are able to find a set of parameters that reproduces the observed radio through infrared quiescent-state SED while respecting the X-ray limits, with the ``(sub)millimeter bump'' naturally captured by emission from the ADAF, while jet emission dominates at lower frequencies.  At radio frequencies, the model image structure decreases in size with increasing frequency, following the same $\text{FWHM} \propto \nu^{-1}$ behavior observed in VLBI observations of \sgra.  Though the jet emission predicted by our model is likely too faint to have been detected in existing VLBI images, it is nevertheless present at observing frequencies of tens of GHz at a level that should be accessible to the ngVLA \citep{Selina_2024}.

There are a number of relevant limitations to our current implementation of the jet model.  Our radiative transfer is carried out assuming flat-space ray propagation (i.e., without light bending), meaning that emission close to the black hole is not properly treated.  We currently compute only the total intensity emission in our numerical implementation, meaning that polarized emission is not modeled.  Furthermore, our model currently predicts only nonthermal synchrotron emission, with no treatment of other emission processes such as inverse Compton emission (likely relevant for describing the high-energy SEDs of some jetted systems) or a thermal electron population (which may be present within the jet or jet sheath).  We note that none of these limitations is inherent to the model itself -- indeed, the original \citet{Tsunetoe_2025} implementation included both general relativistic effects and full-polarization radiative transfer -- but rather specific to our current numerical implementation, and that they could plausibly be relaxed with further development.  A more intrinsic limitation is that our model necessarily describes a steady-state jet, and thus it is fundamentally incapable of predicting any dynamics or variability in these systems.

As high-frequency VLBI telescopes become increasingly capable and provide access to ever-finer angular scales, more astrophysical jet systems will come into view over a wide range of spatial scales.  Indeed, the EHT \citep{EHT_2024}, next-generation EHT \citep{Johnson_2023}, and Black Hole Explorer \citep{Johnson_2024} projects all include jet science among the core motivations for their development.  Our goal with the numerical implementation of the jet model described in this paper is to provide a self-contained Python code that can be used to quickly explore the physical parameters needed to produce the emission observed from these jetted systems.

\software{\texttt{astropy} \citep{Astropy_2013,Astropy_2018,Astropy_2022}, \texttt{ChatGPT} \citep{OpenAI_2025}, \texttt{Claude} \citep{Anthropic_2026}, \texttt{corner} \citep{ForemanMackey_2016}, \texttt{dynesty} \citep{Speagle_2020,Koposov_2023}, \texttt{ehtim} \citep{Chael_2016,Chael_2018,Chael_2023}, \texttt{ehtplot} \citep{Chan_2021}, \texttt{ffjet} \citep{ffjet_zenodo}, \texttt{JAX} \citep{Bradbury_2018,DeepMind_2020}, \texttt{matplotlib} \citep{Hunter_2007}, \texttt{numba} \citep{Lam_2015}, \texttt{numpy} \citep{Harris_2020}, \texttt{NumPyro} \citep{Bingham_2019,Phan_2019}, \texttt{paramsurvey}\footnote{\url{https://github.com/wumpus/paramsurvey}}, \texttt{ray} \citep{Moritz_2017}, \texttt{scikit-learn} \citep{Pedregosa_2011}, \texttt{scipy} \citep{Virtanen_2020}, \texttt{stochastic optics} \citep{Johnson_2016}, \texttt{tqdm} \citep{daCostaLuis_2019}}

\begin{acknowledgments}
We thank Lindy Blackburn, Andrew Chael, Zack Gelles, Michael Johnson, and Pawan Kumar for conversations and discussions that have been helpful throughout the course of this project.  We thank an anonymous referee for constructive comments that improved the quality of this paper.

Support for DWP was provided by the NSF through grants AST-1935980, AST-2034306, and AST-2535855; and by the Gordon and Betty Moore Foundation through grants GBMF5278 and GBMF10423. 
YT is grateful for support from JSPS (Japan Society for the Promotion of Science) Overseas Research Fellowship, the National Natural Science Foundation of China (Grant No. 12325302, 11933007), and the Shanghai Pilot Program for Basic Research, Chinese Academy of Sciences, Shanghai Branch (JCYJ-SHFY-2021-013). 
This work was supported in part by the Black Hole Initiative at Harvard University, which is funded by the John Templeton Foundation (grants 60477, 61479, and 62286) and the Gordon and Betty Moore Foundation (grant GBMF8273).

This research has made use of the NASA/IPAC Extragalactic Database (NED), which is funded by the National Aeronautics and Space Administration and operated by the California Institute of Technology.  This research has made use of the SIMBAD database, operated at CDS, Strasbourg, France.

\end{acknowledgments}

\bibliography{references}{}
\bibliographystyle{aasjournalv7}

\appendix
\numberwithin{equation}{section}

\section{Redshift factor and fluid-frame magnetic field} \label{app:BfieldFluidFrame}

An important angle for synchrotron emission is the photon pitch angle $\theta_B$, i.e., the angle between emitted photon and the magnetic field.  In the observer's frame (and given our simplifying assumption that photons propagate directly towards the observer), we have

\begin{equation}
\cos(\theta_B) = - \frac{B_z}{|\boldsymbol{B}|} . \label{eqn:EmissionAngle}
\end{equation}

\noindent However, the emitting material resides in the fluid frame, and we need to determine $\theta_B$ in the frame comoving with the emitter.

Starting with the photon propagation vector $\hat{\boldsymbol{k}} = (-\sin(i),0,\cos(i))$ in the observer's frame, we can split $\hat{\boldsymbol{k}}$ into components that are parallel and perpendicular to the fluid velocity $\boldsymbol{v}$,

\begin{subequations}
\begin{align}
k_{\parallel} & = \hat{\boldsymbol{k}} \cdot \hat{\boldsymbol{v}} \\
\boldsymbol{k}_{\perp} & = \hat{\boldsymbol{k}} - k_{\parallel} \hat{\boldsymbol{v}} .
\end{align}
\end{subequations}

\noindent Relativistic aberration modifies these components as

\begin{subequations}
\begin{align}
k_{\parallel}' & = \frac{k_{\parallel} - \beta}{1 - \beta k_{\parallel}} \\
\boldsymbol{k}_{\perp}' & = \frac{\boldsymbol{k}_{\perp}}{\gamma_f \big( 1 - \beta k_{\parallel} \big)} ,
\end{align} \label{eqn:AberratedPhotonComponents}
\end{subequations}

\noindent such that

\begin{equation}
\hat{\boldsymbol{k}}' = \frac{\boldsymbol{k}_{\perp}' + k_{\parallel}' \hat{\boldsymbol{v}}}{| \boldsymbol{k}_{\perp}' + k_{\parallel}' \hat{\boldsymbol{v}} |} . \label{eqn:TransformedBField}
\end{equation}

We need to similarly boost the magnetic field.  Given that the fluid-frame electric field is zero-valued after the transformation, we can use

\begin{equation}
\boldsymbol{B}' = \frac{1}{\gamma_f} \boldsymbol{B}_{\perp} + \boldsymbol{B}_{\parallel} , \label{eqn:BFieldFluid}
\end{equation}

\noindent after which the comoving angle is given by

\begin{equation}
\cos(\theta_B') = \frac{\hat{\boldsymbol{k}}' \cdot \boldsymbol{B}'}{|\boldsymbol{B}'|} .
\end{equation}

\noindent Note that we transform both the photon propagation vector and the magnetic field after incorporating the velocity suppression and Lorentz factor cap described in \autoref{sec:FluidVelocity}.

The net redshift factor $g \equiv \nu/\nu'$ will be a combination of the Doppler redshift from the fluid motion and the gravitational redshift, given by

\begin{equation}
g = \frac{\alpha}{\gamma_f \big( 1 - \beta k_{\parallel} \big)} . \label{eqn:RedshiftFactor}
\end{equation}

\noindent Here, $\alpha$ is the lapse defined in \autoref{eqn:Lapse}.

\section{Electron cooling regimes} \label{app:ElectronCoolingRegimes}

As described in \autoref{sec:eDF}, the specific form of the double power-law electron distribution function over Lorentz factor $\gamma'$ depends on the cooling regime.  This cooling regime in turn is determined by whether the relative values of the critical cooling Lorentz factor $\gamma_c'$ (\autoref{eqn:CoolingGamma}) is larger or smaller than the minimum Lorentz factor of the injected power-law electron population $\gamma_m'$.  Electrons in a region of the jet where $\gamma_c' > \gamma_m'$ are in the slow-cooling regime, where the distribution function is given by \autoref{eqn:eDF_gamma} with

\begin{subequations}
\begin{align}
p_1 & = p \\
p_2 & = p+1 \\
\gamma_1' & = \gamma_m' \\
\gamma_2' & = \gamma_c' \\
\gamma_3' & = \gamma_{\text{max}}' .
\end{align}
\end{subequations}

\noindent Electrons in a region of the jet where $\gamma_c' < \gamma_m'$ are in the fast-cooling regime, where the distribution function is given by \autoref{eqn:eDF_gamma} with

\begin{subequations}
\begin{align}
p_1 & = 2 \\
p_2 & = p+1 \\
\gamma_1' & = \gamma_c' \\
\gamma_2' & = \gamma_m' \\
\gamma_3' & = \gamma_{\text{max}}' .
\end{align}
\end{subequations}

\noindent A derivation of the distribution functions for these two cooling regimes is provided in \citet{Tsunetoe_2025}.

\section{Synchrotron emissivity and absorption} \label{app:SynchEmisAbsorp}

Following \citet{Dexter_2016} and \citet{Tsunetoe_2023}, the synchrotron emissivity for Stokes I is given by

\begin{equation}
j_I'(\nu') = \frac{\sqrt{3} n_e' e^3 B' \sin(\theta_B')}{4 \pi m_e c^2} \int_0^{\infty} f(\gamma') F\left( \frac{\nu'}{\nu_c'} \right) d\gamma' , \label{eqn:SynchStokesIEmissivity}
\end{equation}

\noindent where $f(\gamma')$ is the electron distribution function over Lorentz factor, $\theta_B'$ is the angle between the magnetic field and the emission direction in the fluid frame, and

\begin{equation}
\nu_c'(\gamma') = \frac{3 e B' \gamma'^2}{4 \pi m_e c} \sin(\theta_B') \label{eqn:CharacteristicSynchrotronFreq}
\end{equation}

\noindent is a critical frequency at which electrons with Lorentz factor $\gamma'$ preferentially emit.  The $F(x)$ term in the integrand of \autoref{eqn:SynchStokesIEmissivity} is given by the integral of a Bessel function,

\begin{equation}
F(x) = x \int_{x}^{\infty} K_{5/3}(y) dy .
\end{equation}

As in \citet{Tsunetoe_2025}, we compute the synchrotron emissivity corresponding to an anisotropic double power-law distribution of electrons.  Given the form of the distribution in \autoref{eqn:eDF}, the emissivity becomes

\begin{align}
j_I'(\nu') & = \frac{n_e' e^2 \nu_p' \left[ 1 + (\eta - 1) \cos^2(\theta_B') \right]^{-p_{\eta}/2}}{2 \sqrt{3} c A_{\gamma} A_{\xi}} \Bigg[ \left( \frac{\nu'}{\nu_p'} \right)^{-(p_1-1)/2} \Big[ G_I^{(p_1)}(x_2) - G_I^{(p_1)}(x_1) \Big] \nonumber \\
& \qquad + \gamma_2'^{p_2 - p_1} \left( \frac{\nu'}{\nu_p'} \right)^{-(p_2-1)/2} \Big[ G_I^{(p_2)}(x_3) - G_I^{(p_2)}(x_2) \Big] \Bigg] . \label{eqn:Emissivity}
\end{align}

\noindent where $\nu_p' = \nu_c'(\gamma') / \gamma'^2$, $A_{\gamma}$ and $A_{\xi}$ are normalization terms for the distribution function,

\begin{subequations}
\begin{align}
A_{\gamma} & = \frac{\gamma_2'^{1-p_1} - \gamma_1'^{1-p_1}}{1 - p_1} + \gamma_2'^{p_2 - p_1} \left( \frac{\gamma_3'^{1-p_2} - \gamma_2'^{1-p_2}}{1 - p_2} \right) \\
A_{\xi} & = \int_0^1 \Big[ 1 + (\eta - 1) \mu^2 \Big]^{-p_{\eta}/2} d\mu ,
\end{align}
\end{subequations}

\noindent and $G_I^{(p)}(x)$ is given by (see \citealt{Dexter_2016})

\begin{equation}
G_I^{(p)}(x) = \int_x^{\infty} z^{(p-3)/2} F(z) dz .
\end{equation}

\noindent The arguments of the $G_I$ terms in \autoref{eqn:Emissivity} are dimensionless frequencies

\begin{subequations}
\begin{align}
x_1 & = \frac{\nu'}{\nu_c'(\gamma_1')} \\
x_2 & = \frac{\nu'}{\nu_c'(\gamma_2')} \\
x_3 & = \frac{\nu'}{\nu_c'(\gamma_3')} ,
\end{align}
\end{subequations}

\noindent where $\gamma_1'$, $\gamma_2'$, and $\gamma_3'$ are the transition Lorentz factors between the branches in the distribution function (see \autoref{eqn:eDF_gamma}).

We compute the synchrotron absorption coefficient in an analogous manner, starting with the expression from \citet{Dexter_2016}

\begin{equation}
\alpha_I'(\nu') = - \frac{\sqrt{3} n_e' e^3 B' \sin(\theta_B')}{8 \pi m_e^2 c^2 \nu'^2}  \int_0^{\infty} \gamma'^2 \frac{d}{d\gamma'}\left[ \frac{f(\gamma')}{\gamma'^2} \right] F\left( \frac{\nu'}{\nu_c'} \right) d\gamma' . \label{eqn:SynchStokesIAbsorption}
\end{equation}

\noindent Plugging in our anisotropic double power-law distribution function yields

\begin{align}
\alpha_I'(\nu') & = \frac{n_e' e^2 \left[ 1 + (\eta - 1) \cos^2(\theta_B') \right]^{-p_{\eta}/2}}{4 \sqrt{3} m_e c \nu_p' A_{\gamma} A_{\xi}} \Bigg[ (p_1+2) \left( \frac{\nu'}{\nu_p'} \right)^{-(p_1+4)/2} \Big[ Ga_I^{(p_1)}(x_2) - Ga_I^{(p_1)}(x_1) \Big] \nonumber \\
& \qquad + (p_2+2) \gamma_2'^{p_2 - p_1} \left( \frac{\nu'}{\nu_p'} \right)^{-(p_2+4)/2} \Big[ Ga_I^{(p_2)}(x_3) - Ga_I^{(p_2)}(x_2) \Big] \Bigg] , \label{eqn:AbsorptionCoefficient}
\end{align}

\noindent where $Ga_I^{(p)}(x)$ is given by

\begin{equation}
Ga_I^{(p)}(x) = \int_x^{\infty} z^{(p-2)/2} F(z) dz .
\end{equation}

For our numerical implementation of the jet model, we pre-compute the $G_I^{(p)}(x)$ and $Ga_I^{(p)}(x)$ functions and then interpolate as necessary from the tabulated values.

\section{Local electron distribution function} \label{app:LocalEDF}

The broken power-law of \autoref{eqn:eDF_gamma} describes a steady-state distribution function in which electrons are injected locally and assumed to have cooled for a time $t_c'$ (\autoref{eqn:CoolingTime}), with no electrons having Lorentz factor smaller than $\gamma_1'$.  However, the electrons are also advected down the jet, so that each location should also receive the accumulated population of electrons from further upstream \citep[e.g.,][]{Kayama_2022}.  If these transported electrons lost energy only to synchrotron emission, it should result in an accumulation of electrons with energies below the cooling break, in numbers that grow with distance along the jet (although only logarithmically). However, the effect on the spectrum is mitigated by two additional assumptions underpinning our use of the broken power-law eDF described by \autoref{eqn:eDF_gamma}, namely, (1) that the electrons cool adiabatically in addition to their radiative losses, which moves the older electrons to Lorentz factors below $\gamma_1'$ rather than letting them accumulate around $\gamma_1'$; and (2) that electrons with Lorentz factors below $\gamma_1'$ contribute negligibly to the observed emission within the frequency range of interest for this paper (i.e., ${\sim}10^9$-$10^{20}$\,Hz).

\subsection{Adiabatic cooling}

As a fluid element expands, the momentum of each electron within it decreases as $V'^{-1/3}$, where $V' \propto R^2 \gamma_f \beta$ is the comoving volume of the element.  With $R \propto z^s$ and $\gamma_f \beta \propto z^{\varepsilon}$ -- where $\varepsilon \approx s$ when the flow is accelerating and $\varepsilon = 0$ once $\gamma_f = \gamma_{\infty}$ -- the Lorentz factor of a relativistic electron therefore decreases as $z^{-\alpha}$, with

\begin{equation}
\alpha = \frac{2s + \varepsilon}{3} . \label{eqn:AdiabaticIndex}
\end{equation}

\noindent Neglecting synchrotron losses, an electron injected at $z_i$ with Lorentz factor $\gamma_i'$ thus has

\begin{equation}
\gamma'(z) = \gamma_i' \left( \frac{z}{z_i} \right)^{-\alpha} ,
\end{equation}

\noindent such that for $s = 0.6$, we have $0.4 \leq \alpha \leq 0.6$, and an electron loses a factor of $e$ in energy every $1/\alpha \approx 1.7$-$2.5$ logarithmic intervals in distance.

To determine how adiabatic cooling limits the accumulation of electrons, we count the number of electrons $N$ per unit Lorentz factor that the fluid element contains as it moves along the jet.  In each logarithmic interval of distance $d\ln(z_i)$, the fluid element receives a new batch of electrons with the injected spectrum of \autoref{eqn:InjectedeDF},

\begin{equation}
dN_i \propto \gamma_i'^{-p} d\gamma_i' d\ln(z_i) = \gamma'^{-p} \left( \frac{z}{z_i} \right)^{- \alpha (p-1)} d\gamma' d\ln(z_i) ,
\end{equation}

\noindent for $\gamma_i' \geq \gamma_m'$.  Compared with the batch injected locally ($z_i = z$), a batch injected at $z_i$ thus contributes a fraction $(z/z_i)^{-\alpha(p-1)}$ of the electrons at each $\gamma'$.  The population at $z$ is then the sum of these contributions from all upstream injection sites, beginning at some $z_0$ where injection starts,

\begin{equation}
N \propto \gamma'^{-p} \int_{\ln(z_0)}^{\ln(z)} \left( \frac{z}{z_i} \right)^{- \alpha (p-1)} d\ln(z_i) = \frac{\gamma'^{-p}}{\alpha (p-1)} \left[ 1 - \left( \frac{z}{z_0} \right)^{-\alpha(p-1)} \right] ,
\end{equation}

\noindent which becomes

\begin{equation}
N \propto \frac{\gamma'^{-p}}{\alpha (p-1)}
\end{equation}

\noindent for $z \gg z_0$.  Without adiabatic losses ($\alpha = 0$), every batch would contribute in full at every $d\ln(z_i)$, such that the number of electrons with Lorentz factors in the range $(\gamma', \gamma'+d\gamma')$ would grow without limit with $z$ as $\ln(z/z_0)$.  With adiabatic losses, older batches contribute fractionally less and less, so that the population ultimately saturates at $1/[\alpha(p-1)]$ times the locally injected one.

\subsection{Low-energy electrons}

Our assumed eDF \autoref{eqn:eDF_gamma} does not contain any electrons that have cooled below $\gamma_1'$.  An electron with $\gamma' < \gamma_1' \leq \gamma_m'$ radiates primarily below the critical frequency (\autoref{eqn:CharacteristicSynchrotronFreq}) of electrons with $\gamma' = \gamma_m'$,

\begin{equation}
\nu_m' \equiv \nu_c'(\gamma_m') \approx \left( 0.42 \text{\,GHz} \right) \left( \frac{\gamma_m'}{10} \right)^2 \left( \frac{B' \sin(\theta_B')}{1\,\text{G}} \right) , \label{eqn:NuMPrime}
\end{equation}

\noindent above which its emission is exponentially suppressed by the factor $F(x)$ in \autoref{eqn:SynchStokesIEmissivity}.  Observed frequencies will be higher than these fluid-frame frequencies by factors of a few (e.g., $g \approx 1$-$3$ in our fiducial \m87 model) because of Doppler boosting.  Electrons that cool to non-relativistic energies radiate at even lower frequencies near the gyrofrequency.

\section{Projected jet shape} \label{app:ProjectedJetShape}

For a distant observer, the silhouette of the jet will be determined by the set of points where the line-of-sight vector $\hat{\boldsymbol{\ell}}$ is tangent to the outer jet surface (or equivalently, where it is perpendicular to the jet surface normal).  We can rewrite the expression in \autoref{eqn:JetProjRTheta} for the jet edge in Cartesian coordinates to construct the function

\begin{equation}
J(x,y,z) \equiv \left( \frac{x^2 + y^2 + z^2}{r_H^2} \right)^{-(2 - 2s)/2} + \frac{z}{\sqrt{x^2 + y^2 + z^2}} - 1 ,
\end{equation}

\noindent such that $J(x,y,z) = 0$ defines the jet surface.  The normal to the jet surface can then be determined by taking the gradient $\boldsymbol{\nabla} J$,

\begin{equation}
\boldsymbol{\nabla} J = - \frac{1}{r^2} \begin{pmatrix}
x (2 - 2s) \left( \frac{r}{r_H} \right)^{-(2 - 2s)} + x \cos(\theta) \\
y (2 - 2s) \left( \frac{r}{r_H} \right)^{-(2 - 2s)} + y \cos(\theta) \\
z (2 - 2s) \left( \frac{r}{r_H} \right)^{-(2 - 2s)} + z \cos(\theta) - r
\end{pmatrix} ,
\end{equation}

\noindent from which the projected jet shape will be defined by where $\hat{\boldsymbol{\ell}} \cdot \boldsymbol{\nabla} J = 0$.

Given an inclination angle $i$, we can write the line-of-sight vector in Cartesian coordinates as $\hat{\boldsymbol{\ell}} = (\sin(i),0,\cos(i))$, such that the image-plane coordinates $(x',y')$ are given by

\begin{subequations}
\begin{align}
x' & = z \sin(i) - x \cos(i) \\
y' & = y .
\end{align} \label{eqn:XprimeYprime}
\end{subequations}

\noindent Computing the dot product $\hat{\boldsymbol{\ell}} \cdot \boldsymbol{\nabla} J = 0$ and imposing the jet surface condition $J(x,y,z) = 0$ yields (after some algebra) an expression for $\cos(\phi)$ in terms of $\theta$,

\begin{equation}
\cos(\phi) \equiv C(\theta) = \cot(i) \tan\left( \frac{\theta}{2} \right) \left( \frac{1 + (2s - 1) \cos(\theta)}{2 - 2s + (2s - 1) \cos(\theta)} \right) . \label{eqn:JetProjCosPhi}
\end{equation}

\noindent Plugging \autoref{eqn:JetProjRTheta} and \autoref{eqn:JetProjCosPhi} into \autoref{eqn:XprimeYprime} yields parametric expressions for $(x',y')$ in terms of $\theta$,

\begin{subequations}
\begin{align}
x' & = r(\theta) \Big[ \cos(\theta) \sin(i) - \sin(\theta) \cos(i) C(\theta) \Big] \\
y' & = r(\theta) \sin(\theta) \sqrt{1 - \cos^2(\phi)} .
\end{align}
\end{subequations}

\noindent Requiring $|\cos(\phi)| \leq 1$ imposes a maximum value $\theta_{\text{max}}$ such that $\theta$ lives in the range $(0, \theta_{\text{max}}]$; in general, the value of $\theta_{\text{max}}$ needs to be determined numerically.

\section{Longitudinal jet intensity profile} \label{app:LongitudinalProfile}

In the slow-cooling regime when $\nu_m < \nu < \nu_c$, we have

\begin{equation}
j_{\nu} \propto n_e' B \left( \frac{\nu}{\nu_m} \right)^{-(p-1)/2} .
\end{equation}

\noindent Because $\nu_m \propto B$ and $n_e' \propto B^2$, then at a fixed observing frequency $\nu$ we have

\begin{equation}
j_{\nu} \propto B^{(p+5)/2} .
\end{equation}

\noindent At large $r$, the magnetic field strength scales as $B \sim B_{\phi} \propto R^{-1} \propto z^{-s}$, so that

\begin{equation}
j_{\nu} \propto z^{-s(p+5)/2} .
\end{equation}

\noindent The intensity along a line-of-sight through the jet should, in the optically thin limit, scale as $I_{\nu} \propto j_{\nu} R$, and we have $R \propto z^s$.  So we have the expected scaling

\begin{equation}
I_{\nu} \propto z^{-s(p+3)/2} . \label{eqn:LongitudinalProfile}
\end{equation}

\section{\sgra SED measurements} \label{app:SEDMeasurements}

\autoref{tab:SEDMeasurementsSgrA} lists the \sgra SED measurements used for comparison in \autoref{fig:SED_comparison_SgrA}.  The majority of these measurements were initially compiled in \citet{SgrA_Paper2}, and all are meant to capture the quiescent state of \sgra.  We note that recent JWST mid-infrared observations of \sgra \citep{vonFellenberg_2025,Michail_2026} were obtained during a flare and thus are not included in our comparison.

\startlongtable
\begin{deluxetable}{lcccc}
\tablecolumns{5}
\tablecaption{\sgra SED measurements\label{tab:SEDMeasurementsSgrA}}
\tablehead{\colhead{Frequency} & \colhead{Instrument} & \colhead{Angular scale} & \colhead{Flux density} & \colhead{Reference}}
\startdata
$1.16 \times 10^{9}$\,Hz & VLA & 1.5\,arcsec & $0.52 \pm 0.09$\,Jy & \citet{An_2005} \\
$1.3 \times 10^{9}$\,Hz & VLA & 1.3\,arcsec & $0.52 \pm 0.06$\,Jy & \citet{An_2005} \\
$1.5 \times 10^{9}$\,Hz & VLA & 1.15\,arcsec & $0.62 \pm 0.06$\,Jy & \citet{An_2005} \\
$1.6 \times 10^{9}$\,Hz$^{a}$ & VLA & 4\,arcsec & $0.592 \pm 0.028$\,Jy & \citet{Bower_2015} \\
$1.74 \times 10^{9}$\,Hz & VLA & 1\,arcsec & $0.63 \pm 0.05$\,Jy & \citet{An_2005} \\
$3.1 \times 10^{9}$\,Hz$^{a}$ & VLA & 4\,arcsec & $0.702 \pm 0.032$\,Jy & \citet{Bower_2015} \\
$4.86 \times 10^{9}$\,Hz & VLA & 0.34\,arcsec & $0.66 \pm 0.04$\,Jy & \citet{An_2005} \\
$5.4 \times 10^{9}$\,Hz$^{a}$ & VLA & 4\,arcsec & $0.870 \pm 0.118$\,Jy & \citet{Bower_2015} \\
$8.46 \times 10^{9}$\,Hz & VLA & 0.2\,arcsec & $0.69 \pm 0.03$\,Jy & \citet{An_2005} \\
$9 \times 10^{9}$\,Hz$^{a}$ & VLA & 4\,arcsec & $0.932 \pm 0.129$\,Jy & \citet{Bower_2015} \\
$1.4 \times 10^{10}$\,Hz$^{a}$ & VLA & 4\,arcsec & $1.075 \pm 0.135$\,Jy & \citet{Bower_2015} \\
$1.5 \times 10^{10}$\,Hz$^{b}$ & VLA & 2\,arcsec & $0.834 \pm 0.13$\,Jy & \citet{Herrnstein_2004} \\
$2.11 \times 10^{10}$\,Hz$^{a}$ & VLA & 4\,arcsec & $1.164 \pm 0.052$\,Jy & \citet{Bower_2015} \\
$2.2 \times 10^{10}$\,Hz & EAVN & 9\,mas & $1.07 \pm 0.11$\,Jy & \citetalias{SgrA_Paper2} \\
$2.3 \times 10^{10}$\,Hz$^{b}$ & VLA & 2\,arcsec & $0.926 \pm 0.16$\,Jy & \citet{Herrnstein_2004} \\
$3.2 \times 10^{10}$\,Hz$^{a}$ & VLA & 4\,arcsec & $1.382 \pm 0.087$\,Jy & \citet{Bower_2015} \\
$4.09 \times 10^{10}$\,Hz$^{a}$ & VLA & 4\,arcsec & $1.485 \pm 0.073$\,Jy & \citet{Bower_2015} \\
$4.3 \times 10^{10}$\,Hz$^{b}$ & VLA & 2\,arcsec & $1.001 \pm 0.21$\,Jy & \citet{Herrnstein_2004} \\
$4.3 \times 10^{10}$\,Hz & EAVN & 5\,mas & $1.35 \pm 0.14$\,Jy & \citetalias{SgrA_Paper2} \\
$8.7 \times 10^{10}$\,Hz$^{c}$ & GMVA & 3\,mas & $1.9 \pm 0.2$\,Jy & \citetalias{SgrA_Paper2} \\
$9.9 \times 10^{10}$\,Hz$^{c}$ & GMVA & 3\,mas & $2.05 \pm 0.2$\,Jy & \citetalias{SgrA_Paper2} \\
$1.52 \times 10^{11}$\,Hz & IRAM 30m & 15.7\,arcsec & $2.9 \pm 0.25$\,Jy & \citet{Falcke_1998} \\
$2.21 \times 10^{11}$\,Hz & ALMA/SMA & 1\,arcsec & $2.4 \pm 0.2$\,Jy & \citetalias{SgrA_Paper2} \\
$2.26 \times 10^{11}$\,Hz$^{d}$ & ALMA & 6\,arcsec & $3.68 \pm 0.66$\,Jy & \citet{Bower_2015} \\
$2.33 \times 10^{11}$\,Hz$^{e}$ & ALMA & 1.5\,arcsec & $2.886 \pm 0.29$\,Jy & \citet{Bower_2019} \\
$3.48 \times 10^{11}$\,Hz$^{f}$ & ALMA & 6\,arcsec & $3.59 \pm 0.87$\,Jy & \citet{Bower_2015} \\
$4.92 \times 10^{11}$\,Hz & ALMA & 0.54\,arcsec & $3.6 \pm 0.72$\,Jy & \citet{Liu_2016} \\
$6.78 \times 10^{11}$\,Hz$^{e}$ & ALMA & 0.5\,arcsec & $2.183 \pm 0.44$\,Jy & \citet{Bower_2019} \\
$8.68 \times 10^{11}$\,Hz$^{e}$ & ALMA & 0.4\,arcsec & $1.864 \pm 0.37$\,Jy & \citet{Bower_2019} \\
$1.2 \times 10^{12}$\,Hz$^{g}$ & \textit{Herschel}/SPIRE & 18\,arcsec & $>0.5$\,Jy & \citet{Stone_2016} \\
$1.87 \times 10^{12}$\,Hz$^{g}$ & \textit{Herschel}/PACS & 12\,arcsec & $>0.27$\,Jy & \citet{vonFellenberg_2018} \\
$3 \times 10^{12}$\,Hz$^{g}$ & \textit{Herschel}/PACS & 7\,arcsec & $>0.16$\,Jy & \citet{vonFellenberg_2018} \\
$3.49 \times 10^{13}$\,Hz$^{h}$ & VLT/VISIR & 0.3\,arcsec & $<84$\,mJy & \citet{Schodel_2011} \\
$6.25 \times 10^{13}$\,Hz$^{i}$ & VLT/NaCo & 0.15\,arcsec & $3.8 \pm 1.3$\,mJy & \citet{Schodel_2011} \\
$7.89 \times 10^{13}$\,Hz$^{i}$ & VLT/NaCo & 0.12\,arcsec & $5.0 \pm 0.6$\,mJy & \citet{Schodel_2011} \\
$1.36 \times 10^{14}$\,Hz$^{j}$ & VLTI/GRAVITY & 0.06\,arcsec & $1.1^{+2.0}_{-0.7}$\,mJy & \citet{GRAVITY_2020} \\
$1.41 \times 10^{14}$\,Hz$^{k}$ & Keck/NIRC2 & 0.05\,arcsec & $1.0^{+1.3}_{-0.6}$\,mJy & \citet{Weldon_2023} \\
$5.68 \times 10^{17}$\,Hz$^{l}$ & \textit{Chandra}/\textit{NuSTAR} & 1.25\,arcsec & $<1.23 \times 10^{-7}$\,Jy & \citetalias{SgrA_Paper2} \\
$7.84 \times 10^{17}$\,Hz$^{l}$ & \textit{Chandra}/\textit{NuSTAR} & 1.25\,arcsec & $<6.27 \times 10^{-8}$\,Jy & \citetalias{SgrA_Paper2} \\
$1.08 \times 10^{18}$\,Hz$^{l}$ & \textit{Chandra}/\textit{NuSTAR} & 1.25\,arcsec & $<3.09 \times 10^{-8}$\,Jy & \citetalias{SgrA_Paper2} \\
$1.49 \times 10^{18}$\,Hz$^{l}$ & \textit{Chandra}/\textit{NuSTAR} & 1.25\,arcsec & $<1.71 \times 10^{-8}$\,Jy & \citetalias{SgrA_Paper2} \\
$2.06 \times 10^{18}$\,Hz$^{l}$ & \textit{Chandra}/\textit{NuSTAR} & 1.25\,arcsec & $<3.90 \times 10^{-9}$\,Jy & \citetalias{SgrA_Paper2} \\
\enddata
\tablecomments{Measurements of the \sgra SED used in \autoref{fig:SED_comparison_SgrA} and in the SED fit of \autoref{app:SEDfittingSgrA}; the compilation here is restricted to observations that characterize \sgra in a non-flaring state.  The ``angular scale'' column lists the largest angular scale to which each measurement is sensitive.\\
$^a$\citet{Bower_2015} made between 8 and 15 measurements of \sgra's flux density with the VLA over 2012-2014, retaining only baselines longer than 50\,k$\lambda$; the values listed here are the mean and RMS of the measured flux densities.\\
$^b$\citet{Herrnstein_2004} made $>$100 measurements of \sgra's flux density with the VLA over 2000-2003; the values listed here are the mean and RMS of the measured flux densities.\\
$^c$\citet{SgrA_Paper2} report four measurements with the GMVA on 2017 April 3 at 86, 88, 98, and 100\,GHz; we average the 86 and 88\,GHz pair and the 98 and 100\,GHz pair.\\
$^d$Mean and RMS of the eight ALMA epochs of \citet{Bower_2015}, averaged over their four spectral windows at 218-233.8\,GHz.\\
$^e$We have combined the systematic uncertainty estimated by \citet{Bower_2019} (10\% in Band 6, 20\% in Bands 9 and 10) in quadrature with the statistical uncertainty.\\
$^f$Same as $d$, but for the four spectral windows at 341.6-353.6\,GHz.\\
$^g$The listed value represents a variability-based lower limit, which \citet{vonFellenberg_2018} determined as the amplitude of the detected flux density variations, which is taken to bound the time-averaged flux density from below \citep{Stone_2016}.\\
$^h$3$\sigma$ upper limit on the mean 8.6\,$\mu$m emission from \citet{Schodel_2011}, corrected for extinction.\\
$^i$Time-averaged, dereddened flux densities at $M'$ band (4.8\,$\mu$m) and $L'$ band (3.8\,$\mu$m), from \citet{Schodel_2011}.\\
$^j$Median and 16$^{\text{th}}$-84$^{\text{th}}$ percentile interval of the dereddened 2017-2019 \textit{K}-band flux distribution measured by \citet{GRAVITY_2020}.\\
$^k$Median and 16$^{\text{th}}$-84$^{\text{th}}$ percentile interval of the dereddened 2020-2022 Keck $K'$-band flux distribution of \citet{Weldon_2023}.\\
$^l$Quiescent joint \textit{Chandra}/\textit{NuSTAR} spectrum from the 2017 April campaign \citepalias{SgrA_Paper2}.  The X-ray emission within this region is thought to be dominated by extended thermal emission from the accretion flow on scales of ${\sim}10^5\,r_g$, with $\lesssim$10\% attributable to the compact source \citep{Neilsen_2013}; we therefore treat these values as upper limits on the emission from the model.}
\end{deluxetable}
\onecolumngrid

\section{Sgr A* SED parameter survey}

\begin{figure}[t]
    \includegraphics[width=0.90\textwidth]{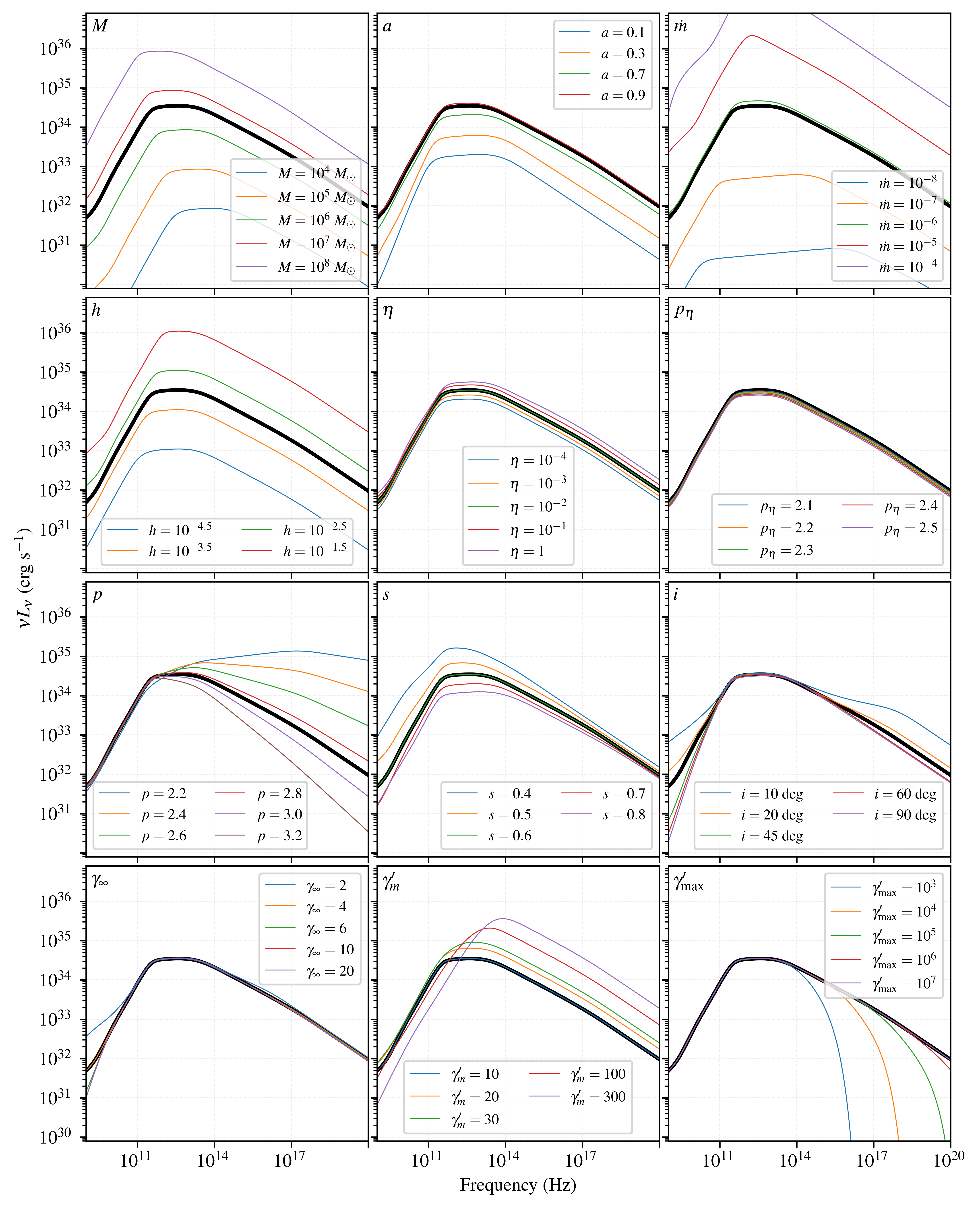}
    \caption{Same as \autoref{fig:parameter_survey}, but showing \sgra rather than \m87.}\label{fig:parameter_survey_SgrA}
\end{figure}

\autoref{fig:parameter_survey_SgrA} shows a gallery of SEDs produced via surveying across each model parameter individually, using the parameters for \sgra in \autoref{tab:Parameters} as our baseline model.  Most of the scalings detailed in \autoref{sec:SED} apply just as well for \sgra as they do for \m87.  The primary exception is the inclination angle $i$, the modification of which causes qualitatively different changes to the \sgra SED compared to the \m87 case.

The two main drivers of the different \sgra SED behavior are its lower accretion rate $\dot{m}$ and higher eDF power-law index $p$ than \m87.  The lower $\dot{m}$ results in the \sgra jet being entirely in the slow-cooling regime, for which $\gamma_c' > \gamma_m'$.  Because $B \propto z^{-s}$, the characteristic Lorentz factor goes as $\gamma' \propto \nu^{1/2} B^{-1/2} \propto \nu^{1/2} z^{s/2}$; i.e., the characteristic Lorentz factor grows with frequency and radius.  In the slow-cooling regime, the spectral dependence is

\begin{equation}
F_{\nu} \propto \left( \frac{\nu}{\nu_m} \right)^{-(p-1)/2}
\end{equation}

\noindent for $\nu_m < \nu < \nu_c$ (see \autoref{eqn:ScalingSPN}).  Because $\gamma' / \gamma_m'$ (and thus $\nu / \nu_m$) increases towards larger radii, the emission at these radii is suppressed by a factor $(\nu / \nu_m)^{0.94}$ for $p = 2.88$.  So for a broad range of frequencies, the \sgra SED is completely dominated by emission from the innermost radii (see \autoref{fig:SED_comparison_SgrA}), where the effects of changing the inclination angle are minimized.  This insensitivity to changing $i$ is evident in \autoref{fig:parameter_survey_SgrA} for frequencies spanning ${\gtrsim}10^{11}$\,Hz.  It remains the case that contributions from the outer jet are most important at lower frequencies, where the emission is more optically thick.  We thus see in \autoref{fig:parameter_survey_SgrA} that the effects of changing the inclination angle -- which preferentially impact the emission at large radii -- are more pronounced at low frequencies than they are for the higher frequencies where the SED is dominated by emission from small $r$.

\section{M87 SED fitting} \label{app:SEDfitting}

In this section we describe the procedure we use to fit the jet model to the broadband \m87 SED observations.  This fitting serves both to demonstrate that a single parameter set can quantitatively reproduce the observed SED across all frequencies and spatial scales simultaneously, as well as to determine the actual values of the model parameters for \m87.

\subsection{Dataset}

We fit the \m87 SED measurements from \citet{M87_MWL}, which consist of 45 flux density measurements at 41 distinct frequencies spanning 1.7\,GHz to $1.2 \times 10^{19}$\,Hz ($\sim$49\,keV).  Each measurement is sensitive to emission on a particular angular scale $\theta_i$, ranging from $\sim$20\,\uas ($\sim$5\,$r_g/D$) for the EHT observations to $\sim$3\,arcseconds (${\sim}10^6$\,$r_g/D$) for Swift observations.  We take the model prediction for measurement $i$ to be the jet emission enclosed within the observed angular scale, which we evaluate by integrating over the model image within the radius $r_i$ given (in units of the gravitational radius $r_g$) by

\begin{equation}
r_i = \frac{\theta_i D}{r_g} , \label{eqn:aperture}
\end{equation}

\noindent where $D = 16.8$\,Mpc is the distance to \m87.  We include an emission contribution from the ADAF model of \citet{Pesce_2021} evaluated at the same $M$, $\dot{m}$, and outer radius $r_i$ as the jet model.  We note that because $r_i$ is measured in units of the gravitational radius, it necessarily changes with the mass of the black hole, and so the physical radius out to which emission is integrated for each measurement is part of the fit.

We adopt a 10\% systematic uncertainty in each SED measurement, to account for the intrinsic variability of \m87 and the uncertain absolute flux density scales of the measurements.  This systematic uncertainty is added in quadrature to the formal measurement uncertainties reported in \citet{M87_MWL}, and it is the dominant uncertainty component for 25 of the 45 measurements.

\subsection{Likelihood and priors}

We assume that all measurements are statistically independent, and we adopt a Gaussian likelihood,

\begin{equation}
\ln(\mathcal{L}) = -\frac{1}{2} \sum_{i} \left( \frac{F_{\text{obs},i} - F_{\text{mod},i}}{\sigma_i} \right)^2 .
\end{equation}

\noindent We fit for eleven parameters -- $M$, $\dot{m}$, $a$, $i$, $s$, $p$, $h$, $\eta$, $\gamma_{\infty}$, $\gamma_m'$, and $\gamma_{\rm{max}}'$ -- and we hold the $p_{\eta}$ parameter fixed at $2$.  The prior distribution for each parameter is specified in \autoref{tab:FittedParameters}.  For the black hole mass prior we use a Gaussian distribution matching the EHT measurement of $M = (6.5 \pm 0.9) \times 10^9$\,\msun \citep{M87Paper6}, and for the accretion rate prior we use a log-uniform distribution spanning the range of model values consistent with the EHT polarimetric constraints \citep{M87Paper7}.  For the inclination angle we adopt an isotropic prior, i.e., uniform in $\cos(i)$.  All remaining parameters have uninformative priors, uniform in either linear or log.

\subsection{Sampling}

We sample the posterior distribution using the No-U-Turn Sampler \citep[NUTS;][]{Hoffman_2014} as implemented in \texttt{NumPyro} \citep{Bingham_2019,Phan_2019}, with twelve chains of 4000 samples each following 2000 warm-up steps.  Each chain is initialized randomly from within the prior distribution, and we assess convergence using the Gelman-Rubin ``split $\hat{R}$'' statistic \citep{GelmanRubin1992}.  Our sampling achieves $\hat{R} \leq 1.006$ for every parameter, with a smallest effective sample size of $\sim$2000.

\subsection{Results}

\begin{figure*}[t]
    \includegraphics[width=1.00\textwidth]{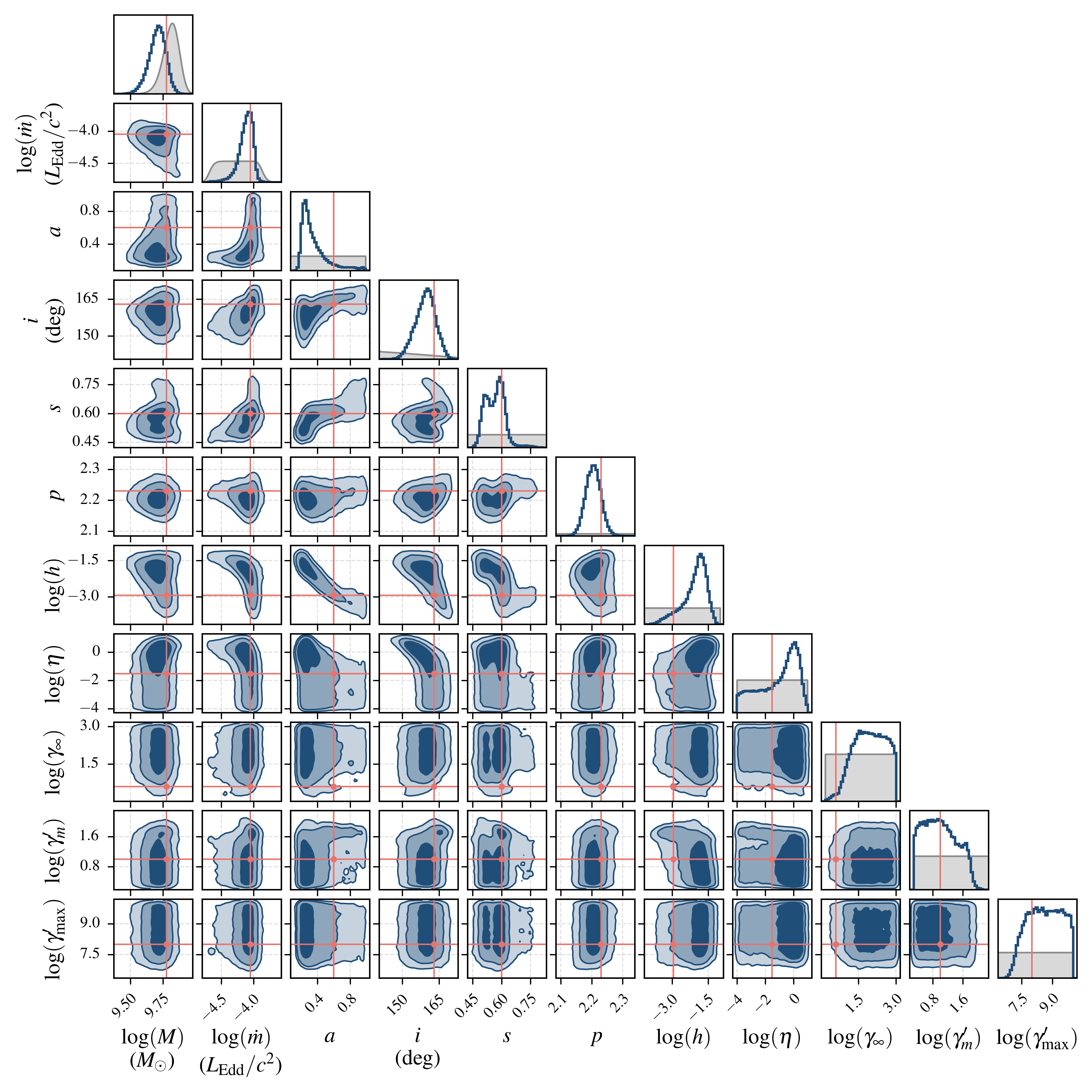}
    \caption{Corner plot showing the 2D and 1D marginal posterior distributions of the jet model parameters fitted to the \m87 SED measurements (\autoref{app:SEDfitting}).  In the 2D marginal panels, the plotted contours enclose 50\%, 90\%, and 99\% of the posterior probability.  The gray histograms in the 1D marginal panels denote the prior distribution, which is independent for each parameter.  The red lines mark our fiducial \m87 solution.}\label{fig:posterior_M87}
\end{figure*}

\autoref{tab:FittedParameters} lists the best-fit values for each parameter, quoted as the median and (16$^{\text{th}}$, 84$^{\text{th}}$) percentiles or as a 99\% one-sided bound for the parameters that the fit constrains from only one side.  1D and 2D marginals of the posterior distribution are plotted in \autoref{fig:posterior_M87}.  The best-fit model achieves $\chi^2 = 33.5$ for 45 measurements and eleven free parameters, corresponding to a reduced $\chi_{\nu}^2 = 0.99$.  I.e., the model is capable of providing a quantitatively good fit to the broadband \m87 SED.

\begin{deluxetable}{lc|cc|cc}
\tablecaption{Best-fit model parameters derived from fitting \m87 and \sgra SEDs
\label{tab:FittedParameters}}
\tablehead{
    \colhead{Parameter} &
    \colhead{Units} &
    \colhead{\m87 prior} &
    \colhead{\m87 fit} &
    \colhead{\sgra prior} &
    \colhead{\sgra fit}
}
\startdata
$M$ & M$_{\odot}$ & $\mathcal{N}(6.5, 0.9) \times 10^9$ & $5.12^{+0.73}_{-0.74} \times 10^9$ & $\mathcal{N}(4.14, 0.08) \times 10^{6}$ & $4.10 \pm 0.09 \times 10^{6}$ \\
$a$ & unitless & $\mathcal{U}(0.01,0.99)$ & $0.304^{+0.189}_{-0.076}$ & $\mathcal{U}(0.01,0.99)$ & $0.618^{+0.285}_{-0.309}$ \\
$i$ & degrees & $\mathcal{U}_{\cos}(90,180)$ & $159.7^{+3.7}_{-4.4}$ & $\mathcal{U}_{\cos}(0,90)$ & $30.7^{+18.7}_{-4.8}$ \\
$\dot{M}$ & M$_{\odot}$\,yr$^{-1}$ & $\mathcal{U}_{\text{log}}(3, 20) \times 10^{-4}$ & $8.63^{+2.63}_{-2.04} \times 10^{-4}$ & $\mathcal{U}_{\text{log}}(10^{-10}, 10^{-6})$ & $8.05^{+3.86}_{-0.93} \times 10^{-9}$ \\
$s$ & unitless & $\mathcal{U}(0.3,0.9)$ & $0.569^{+0.041}_{-0.057}$ & $\mathcal{U}(0.3,0.99)$ & $0.517^{+0.095}_{-0.096}$ \\
$p$ & unitless & $\mathcal{U}(2.02,4.0)$ & $2.205 \pm 0.024$ & $\mathcal{U}(2.02,6.0)$ & $2.811^{+0.162}_{-0.154}$ \\
$\log(h)$ & unitless & $\mathcal{U}(-5, -1)$ & $-1.950^{+0.337}_{-0.628}$ & fixed & $-3$ \\
$\log(\eta)$ & unitless & $\mathcal{U}(-4, 1)$ & $-0.669^{+0.921}_{-2.095}$ & fixed & $-2$ \\
$p_{\eta}$ & unitless & fixed & 2 & fixed & 2 \\
$\log(\gamma_{\infty})$ & unitless & $\mathcal{U}(\log(1.5), 3)$ & $>0.551$ & fixed & $\log(4)$ \\
$\log(\gamma_m')$ & unitless & $\mathcal{U}(\log(2), 3)$ & $<1.862$ & fixed & 1 \\
$\log(\gamma_{\text{max}}')$ & unitless & $\mathcal{U}(3, 10)$ & $>7.070$ & fixed & 8 \\
\enddata
\tablecomments{Prior distributions and posterior constraints from the SED fits described in \autoref{app:SEDfitting} and \autoref{app:SEDfittingSgrA}.  Fitted values for most parameters are quoted as the posterior median with (16$^{\text{th}}$, 84$^{\text{th}}$) percentile intervals.  For $\log(\gamma_{\infty})$, $\log(\gamma_m')$, and $\log(\gamma_{\text{max}}')$ in the \m87 fit, we instead quote a 99\% limit, as the posterior for these parameters is bounded only from one side.  The $\log(\eta)$ parameter is effectively unconstrained by the \m87 SED fit, with substantial posterior support across the full prior range.  $\mathcal{N}(\mu, \sigma)$ denotes a Gaussian prior with mean $\mu$ and standard deviation $\sigma$, $\mathcal{U}(a,b)$ denotes a uniform prior on the interval $[a,b]$, $\mathcal{U}_{\text{log}}(a,b)$ denotes a log-uniform prior on the interval $(a,b)$, and $\mathcal{U}_{\cos}(a, b)$ denotes a uniform prior in $\cos(i)$ for $i$ in $[a, b]$ degrees (corresponding to an isotropic prior on the inclination angle).  Parameters marked as ``fixed'' are held at the listed values.}
\end{deluxetable}

\null

\section{Sgr A* SED fitting} \label{app:SEDfittingSgrA}

This section is analogous to \autoref{app:SEDfitting}, but describes the fitting process for \sgra rather than \m87.

\subsection{Dataset}

We fit the \sgra SED measurements listed in \autoref{tab:SEDMeasurementsSgrA}, which consist of 42 flux density measurements at 41 distinct frequencies spanning 1.16\,GHz to $2.1 \times 10^{18}$\,Hz ($\sim$8.5\,keV), capturing the quiescent emission state of \sgra.  The angular scales range from ${\sim}3$\,mas (${\sim}600$\,$r_g/D$) for the EAVN and GMVA observations to $\sim$18\,arcseconds (${\sim}4 \times 10^6$\,$r_g/D$) for the \textit{Herschel} observations, and we again take the model prediction for each measurement to be the jet emission enclosed within the observed angular scale.  We use $D = 8.15$\,kpc as the distance to \sgra \citep{Reid_2019}, and we again include an emission contribution from the ADAF model of \citet{Pesce_2021} evaluated at the same $M$, $\dot{m}$, and outer radius $r_i$ as the jet model.

Because \sgra is known to be highly time-variable with a degree of variation that increases with frequency through the infrared, we use a frequency-dependent systematic uncertainty rather than the flat 10\% that we adopt for \m87.  Specifically, we assume a systematic uncertainty of 15\% for $\nu \leq 50$\,GHz, 25\% for $50\text{\,GHz} < \nu \leq 300$\,GHz, 30\% for $\nu > 300$\,GHz through the infrared, and then 10\% in the X-ray.

\subsection{Likelihood and priors}

We adopt the same Gaussian likelihood as used in the \m87 fit for most of the SED measurements.  For the two near-infrared measurements -- which represent medians of broad, approximately log-normal flux distributions \citep{GRAVITY_2020,Weldon_2023} -- we instead adopt a likelihood that is Gaussian in $\log(F_{\nu})$ using the widths listed in \autoref{tab:SEDMeasurementsSgrA} (i.e., 0.43 and 0.36\,dex).  We treat the six upper limits (i.e., the one at 8.6\,$\mu$m and the five X-ray measurements) and the three lower limits (i.e., the \textit{Herschel} variability-based measurements) using one-sided Gaussian likelihood terms, which contribute to the total likelihood only when the model violates the limits.

For the \sgra fit, we explore a 6-parameter subset of the 11-parameter space that we fit for \m87.  The six parameters we fit are $M$, $\dot{m}$, $a$, $i$, $s$, and $p$, with the other five parameters held fixed; we continue to hold the twelfth parameter, $p_{\eta}$, fixed at $2$.  For the black hole mass prior we use a Gaussian distribution with mean and standard deviation $M = (4.14 \pm 0.08) \times 10^6$\,\msun, falling between the measurements of \citet{Do_2019} and \citet{GRAVITY_2019}.  We again adopt an isotropic prior on the inclination angle, i.e., uniform in $\cos(i)$ in the range $(0,90)$\,degrees.  We use broad uninformative priors for the other four parameters, uniform in either linear or log.  The prior distribution for each parameter is specified in \autoref{tab:FittedParameters}.

The primary difference between our \sgra fit and our \m87 fit is that for \sgra, we hold the remaining five parameters fixed at values of $h = 10^{-3}$, $\eta = 0.01$, $\gamma_{\infty} = 4$, $\gamma_m' = 10$, and $\gamma_{\text{max}}' = 10^8$.  These parameters are poorly constrained by the SED data for both \m87 and \sgra, but leaving them as free parameters in the \sgra case drives the posterior bulk to a region of parameter space where $h \gg 0.1$ and $s > 0.9$; i.e., when all parameters are free to vary, the \sgra SED prefers a solution that would be physically interpreted as a nearly isotropic outflow in which nonthermal electrons absorb effectively all of the available jet power.  And unlike for \m87, we have no spatially resolved images of any jet in \sgra that we can use to subselect a more obviously jetted configuration from within the broad posterior volume.  We therefore instead opt to use the \m87 fiducial parameter set to pin the least well-constrained of the \sgra model parameters, ensuring that our fit addresses the relevant science question of whether a collimated jet can reproduce the \sgra SED.  We note, however, that the space-filling solution would produce a very low-brightness jet that is difficult to detect in VLBI images of \sgra, which is consistent with the non-detection of such a jet to date.

\subsection{Sampling}

We sample the \sgra posterior via the same strategy that we employ for the \m87 fit, using twelve NUTS chains of 4000 samples each following 2000 warm-up steps and initialized randomly.  Our sampling achieves $\hat{R} \leq 1.007$ for every parameter, with a smallest effective sample size of ${\sim}900$.

\subsection{Results}

\begin{figure*}[t]
    \includegraphics[width=1.00\textwidth]{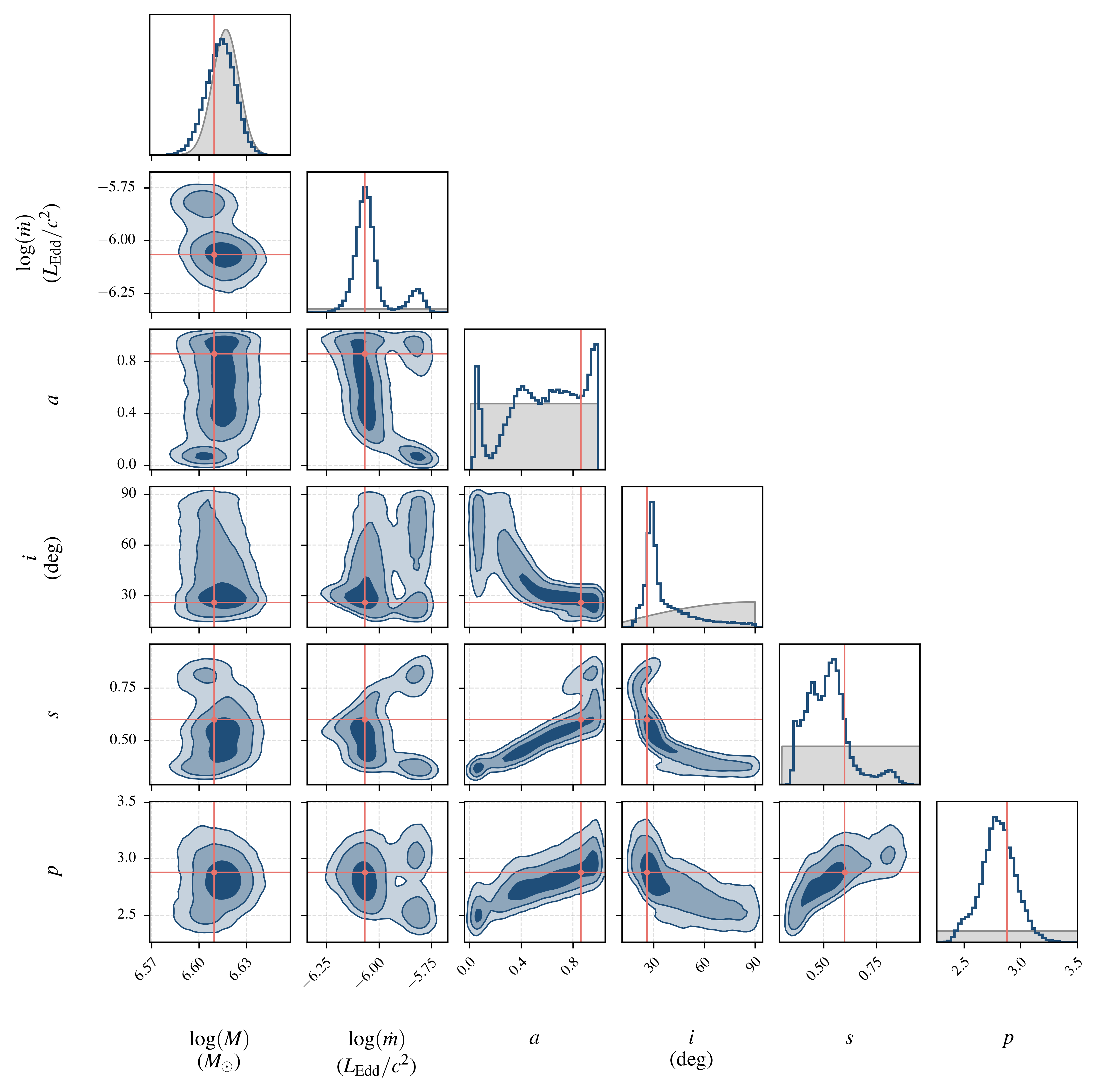}
    \caption{Same as \autoref{fig:posterior_M87}, but for the \sgra SED fit, which explored only 6 of the 11 parameters (\autoref{app:SEDfittingSgrA}).  The red lines denote the fiducial parameter set.} \label{fig:posterior_SgrA}
\end{figure*}

As with the \m87 fit, \autoref{tab:FittedParameters} lists the best-fit values for each parameter in our \sgra fit, with 1D and 2D marginals of the posterior distribution plotted in \autoref{fig:posterior_SgrA}.  The best-fit model achieves $\chi^2 = 34.3$ for 42 measurements, 33 of which have two-sided uncertainties (i.e., they are not one-sided limits) and which are thus the primary $\chi^2$ drivers; with six free parameters, the corresponding reduced $\chi_{\nu}^2 = 1.27$.  I.e., the model is capable of providing a quantitatively good fit to the broadband \sgra SED.

\end{document}